\documentclass[preprint,amsmath,amssymb,aps,pra]{revtex4-2}

\usepackage{graphicx}
\usepackage{dcolumn}
\usepackage{bm}
\usepackage[dvipsnames]{xcolor}
\usepackage{soul}
\usepackage[version=3]{mhchem}
\usepackage{mathrsfs}
\usepackage{bigstrut}

\begin{document}

\title{Linking structure and optical properties of plasmonic nanoparticles on tunable spherical surfaces}

\author{Francesco Brasili$^{1,2}$}
\thanks{These authors contributed equally to this work} 
\author{Angela Capocefalo$^3$}
\thanks{These authors contributed equally to this work}
\author{Giovanni Del Monte$^{1,4}$}
\author{Rodrigo Rivas-Barbosa$^2$}
\author{Javier P\'erez$^5$}
\author{Edouard Chauveau$^6$}
\author{Federico Bordi$^2$}
\author{Carlo Rizza$^3$}
\author{Domenico Truzzolillo$^6$}
\author{Emanuela Zaccarelli$^{1,2}$}
\email{emanuela.zaccarelli@cnr.it}
\author{Simona Sennato$^{1,2}$\vspace{5mm}}
\email{simona.sennato@cnr.it}

\affiliation{\scriptsize $^1$Institute for Complex Systems, National Research Council, Piazzale Aldo Moro 5, 00185, Rome, Italy}
\affiliation{$^2$Department of Physics, Sapienza University of Rome, Piazzale Aldo Moro 5, 00185, Rome, Italy}
\affiliation{$^3$Department of Physical and Chemical Sciences, University of L’Aquila, Via Vetoio, Coppito, 67100, L’Aquila, Italy}
\affiliation{$^4$Soft Condensed Matter and Biophysics, Debye Institute for Nanomaterials Science, Utrecht University, Princetonplein 1, 3584 CC, Utrecht, The Netherlands}
\affiliation{$^5$Synchrotron SOLEIL, L’Orme des Merisiers, D\'epartementale 128, 91190 Saint-Aubin, France}
\affiliation{$^6$Laboratoire Charles Coulomb, UMR 5221, CNRS--Universit\'e de Montpellier, 34095, Montpellier, France}
\affiliation{   }

\begin{abstract}
The complexation of plasmonic nanoparticles (NPs) and thermoresponsive microgels is widely exploited for applications, but a microscopic description of the mechanisms governing the spatial organization of the NPs is still lacking. Combining small angle X-ray scattering, state-of-the-art simulations and a simple toy model, we uncover how the volume phase transition of microgels controls NP-NP interactions, establishing for the first time a microscopic link between plasmon coupling and NP local structure. Our study paves the way to experimentally investigate phase transitions on controlled curved surfaces at the nanoscale.
\end{abstract}

\maketitle 

\newpage
The study of particle systems confined to curved two-dimensional interfaces is a fundamental question in physics, dating back to the classical Thomson problem \cite{thomson1904}. Its straightforward generalization to various natural and synthetic systems, including viral capsids, biological membranes, two-dimensional crystals, colloidosomes, and advanced nanomaterials for photonic and electronic applications, further broaden its relevance \cite{bowick2009,vitelli2006,martin2021,fantoni2012}.
While extensive theoretical studies have explored the equilibrium configuration and thermodynamics of particles on curved surfaces \cite{bozic2019,javidpour2021,meyra2019,carenza2022,viveros2008}, experimental investigations remain scarce and are completely absent at the nanoscale \cite{bausch2003,sun2025,singh2022,guerra2018}. 
Addressing this gap is essential, as inter-particle interactions at the nanoscale deviate from classical additivity \cite{silvera2015}, potentially giving rise to unexpected behaviors distinct from those observed at the microscale and predicted by current numerical models.
To tackle this challenge, we investigate an experimental model system where the spherical constraint is provided by soft polymeric colloids, specifically thermoresponsive microgels. 
Here, we leverage their volume phase transition (VPT), a reversible collapse triggered by temperature increase \cite{fernandez2011}, to dynamically tune the curvature of their external surface, onto which nanoparticles (NPs) are electrostatically adsorbed.
However, the sub-wavelength characteristic dimensions of the microgel--NPs complexes prevent direct imaging via optical microscopy, that has hitherto been the standard method for studying colloidal particles at the curved interface of Pickering emulsions \cite{kelleher2017}. To overcome this limitation, we combine Small Angle X-ray Scattering (SAXS) experiments with molecular dynamics simulations, enabling us to rationalize the temperature-dependent arrangement of adsorbed NPs. As a proof of concept, we focus on a plasmonic system, using gold NPs as interacting colloids to have an additional local probe to monitor interparticle interactions.
In this respect, a pioneering contribution was put forward by Gawlitza \textit{et al.} \cite{gawlitza2013}, who recognized the effect of  the microgel structure on NP loading and on plasmon coupling, later extensively exploited to engineer responsive photonic nanomaterials \cite{suzuki2014,choe2018,sabadasch2020,arif2021,diehl2022}. However, a microscopic understanding of NPs rearrangement and a quantitative connection between their structure and the resulting optical properties have not been established so far.

\begin{figure*}
\includegraphics[width=0.98\textwidth]{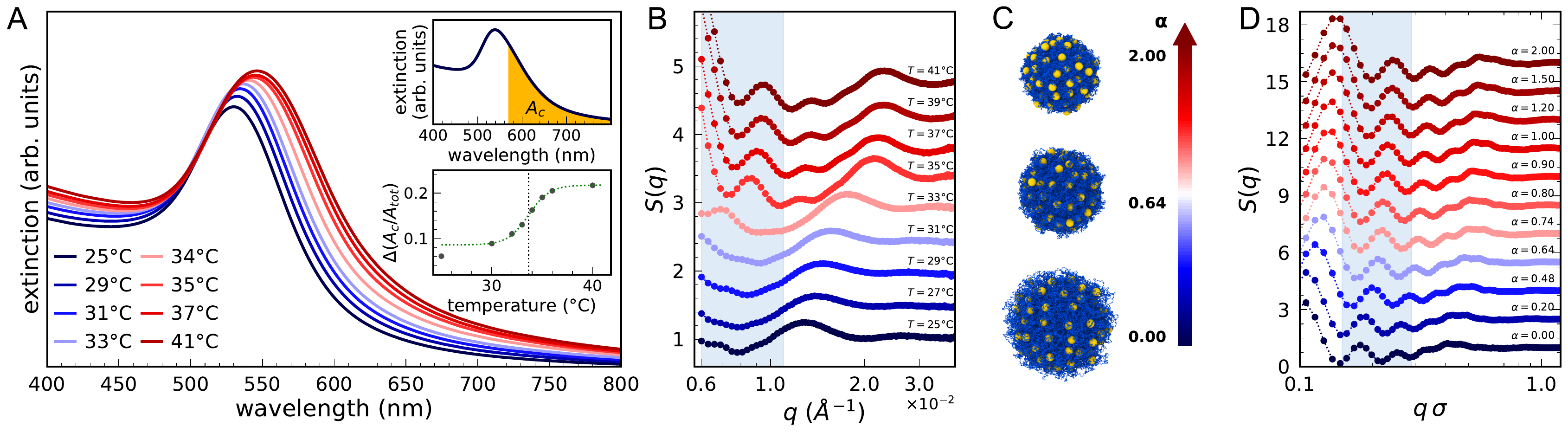}
\caption{\label{fig:1} (A) Optical properties of the microgel--NPs complexes: extinction spectra for different temperatures $T$ for $n=150$; top inset, region of coupled plasmon modes (orange, $A_c$) used to define the degree of coupling $\Delta (A_C/A_{tot})$ (Eq.~S2, details in SM); bottom inset, $\Delta (A_C/A_{tot})$ as a function of $T$ and sigmoidal fit (dashed line) yielding a critical temperature $T_C=33.7\pm0.2$°C. (B) Experimental structure factors of NPs adsorbed onto microgels at different $T$. 
(C) Simulation snapshots for three representative values of the effective temperature $\alpha$; blue and yellow particles represent microgel monomers and NPs, respectively. (D) Numerical $S(q)$ of NPs adsorbed onto a microgel with $N=112k$ and $f=0.02$ for $n=150$ and different $\alpha$-values. Curves in panels (B) and (D) are vertically shifted for clarity; the light blue color highlights the peak mostly described in the text.}
\end{figure*}

In this work, we fill this gap by investigating the temperature-dependent arrangement of adsorbed gold NPs and linking, for the first time, their structural organization to their optical response via extinction spectroscopy and electromagnetic full-wave simulations. We find that, as temperature increases, the NPs approach each other due to microgel shrinkage but do not form clusters. Instead, they rearrange to maximize their geodesic distance, minimizing electrostatic repulsion. To interpret these findings, we introduce a model describing NPs confined to a spherical shell, which accurately captures the structure factors of the NPs within the microgel corona. Furthermore, by incorporating this NP arrangement into electromagnetic simulations, we successfully reproduce the experimental extinction spectra. Our approach establishes a direct relationship between plasmon coupling strength and NP separation, independently of the total number of NPs, laying the groundwork for precise control over the optical properties of soft plasmonic complexes.

We employ cationic poly(N-isopropylacrylamide) (pNIPAM) microgels interacting with anionic, spherical gold NPs and analyze the samples by extinction spectroscopy and SAXS for different temperatures $T$ across the VPT (see experimental methods in Supplemental Material, SM). 
Extinction spectra, reported in Fig.~\ref{fig:1}A, for NP/microgel number ratios $n=150$, show that the LSPR evolves with increasing $T$, exhibiting a redshift and a slight quenching as well as an increase in the extinction at larger wavelengths. Similar results are found for $n=300$ (Fig.~S1). These spectral changes are the signature of plasmon coupling, that is the activation of low-energy collective modes, arising from plasmon hybridization \cite{halas2011} occurring when, driven by the microgel deswelling (NPs alone do not aggregate with $T$, as shown in Fig.~S1), NPs approach each other reaching surface-to-surface distances $d$ of a few nanometers (see SM for details).
We quantify the degree of coupling by the spectral weight $\Delta (A_C/A_{tot})$ of the region between 570 and 800 nm (Fig.~\ref{fig:1}A), relevant for colorimetric detection or surface-enhanced spectroscopy \cite{li2020,capocefalo2022,liu2014,caprara2020}, as defined in Eq.~S2.
Notably, as shown in the inset of Fig.~\ref{fig:1}A, the degree of coupling follows a sigmoidal trend in temperature with an inflection point occurring at $T_C$, demonstrating the possibility to detect the VPT of the microgels through the coupling of the adsorbed NPs.

To gain knowledge on the microscopic NP--NP organization, we measure NP--NP structure factors $S(q)$ (defined in Eq.~S1) as a function of $T$, as reported in Fig.~\ref{fig:1}B. 
Since the measured SAXS intensities solely arise from NPs, due to the much higher contrast of gold with respect to the polymer \cite{suzuki2014}, $S(q)$'s are directly obtained by dividing each scattering curve by the form factor of NPs (Fig.~S2). The complex structure of the soft assemblies makes  the interpretation of the curves and their $T$-dependent evolution far from trivial within the analyzed $q$-range. In particular, the onset of a new peak at $\sim 1 \times 10^{-2} $\AA$^{-1}$ (highlighted in light blue) when temperature exceeds $T_C=33.3\pm0.3$ °C (swelling curve in Fig.~S3) and the concurrent shift of the large band from $1.2 \times 10^{-2} $\AA$^{-1}$ to $2.2 \times 10^{-2} $\AA$^{-1}$ point to a nanoscale rearrangement of NPs across the VPT of the microgels.

To decipher these observations, we resort to molecular dynamics simulations of a well-established microgel model \cite{gnan2017}, that provides a realistic description across the VPT \cite{ninarello2019} also in presence of charged monomers \cite{del2019} and upon interactions with charged NPs \cite{brasili2023} (see SM for details). In the simulations, the increase of temperature is reproduced by the use of a solvophobic potential between the monomers (Eq.~S4), which implicitly accounts for monomer–solvent interaction through a parameter $\alpha$ that plays the role of an effective temperature \cite{soddemann2001}. Good solvent conditions correspond to $\alpha=0$, while the affinity to the solvent gets worse as $\alpha$ increases.
Simulations are conducted for single microgels, made of either $N=14k$  or $N=112k$ monomers of diameter $\sigma$, with a varying fraction $f$ of charged monomers located on the external shell~\cite{del2019}, for values $\alpha$ in the presence of varying amounts of NPs and relative counterions (see SM). 
The  swelling curve of the microgel (Fig.~S4) displays the occurrence of the VPT at $\alpha_C \sim 0.64$.
We analyze microgel--NPs complexes (snapshots of Fig.~\ref{fig:1}C) to calculate NP--NP structure factors, reported in Fig.~\ref{fig:1}D for $N=112k$ and $n=150$, which exhibit very similar features to experiments: high oscillations at low $q$ are followed by smaller peaks at intermediate $q$, showing overall a general shift of the peaks towards higher $q$ with increasing temperature. However, the second peak, which in the simulations is around $0.2\,\sigma^{-1}$ (highlighted in light blue), is already present below the VPT, in contrast to experiments. This result is found in all performed simulations, either varying $f$ and $n$ or modifying the microgel size (Fig.~S5).
\begin{figure*}[t]
\includegraphics[width=0.98\textwidth]{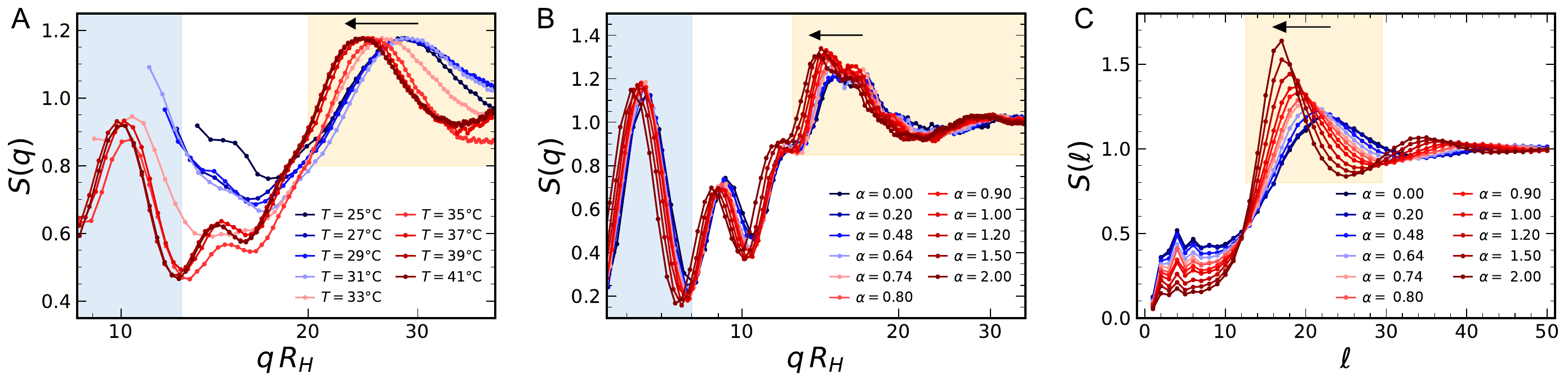}
\caption{\label{fig:2} Structural analysis of NPs adsorbed to microgels across the VPT: $S(q)$ as a function of $q\,R_H$ for SAXS experiments (A) and numerical simulations (B); (C) numerical spherical structure factors $S(\ell)$ of NPs for a microgel with $N=14k$ and $f=0.161$. The shift towards low $q$ or $\ell$ is highlighted by a yellow region; the light blue region is the same as in Fig. 1.} 
\end{figure*}

To shed light on this, we analyze the shift of all peaks towards large $q$ values, filtering out the effect of the microgel deswelling due to the VPT. We thus plot $S(q)$ as a function of $q\,R_H(T)$ in Fig.~\ref{fig:2}A and B, for experiments and simulations, respectively, with $R_H(T)$ being the hydrodynamic radius at each temperature. In this way, it becomes evident that the region $q\,R_H\leq 11$, where the curves above the VPT have a peak, falls outside the experimental $q$-range below the VPT. Therefore, the peak should be already present at low temperatures, similarly to what found in simulations, but its apparent onset at high $T$ is simply due to the fixed experimental $q$-window.
Having established this, we take a closer look to Fig.~\ref{fig:2} where, for both experimental and simulation data, the rescaled curves do not perfectly overlap at different temperatures. Indeed, a shift towards lower $q$ values is observed in both panels, pointing to a variation in NP-NP arrangement as a function of $T$. This effect needs to be appropriately interpreted within the spherical geometry constraint imposed on the NPs by the presence of the microgel. Hence, in order to decouple the contribution of the confining geometry to $S(q)$, we calculate the spherical structure factor $S(\ell)$ which is defined on the surface of a unit sphere as \cite{bozic2019}
\begin{equation}
S(\ell) = 1 + \frac{2}{n} \sum_{(i,j)} P_\ell(\cos\gamma_{ij})
\end{equation}
where the sum is performed over all $(i,j)$ NPs located at $\vec{r}_i$ and  $\vec{r}_j$, with $\gamma_{ij} = 2\arcsin(\frac{1}{2} |\hat{r}_i-\hat{r}_j| )$ 
being the geodesic distance (angular distance on the unit sphere \cite{bozic2019}) between the two NPs, $\hat{r}_{i}$ the versor of $\vec{r}_{i}$ and $P(\ell)$ is the Legendre polynomial of degree $\ell$.
Here, $\ell$ plays an equivalent role to the wave vector $q$ in ordinary space, with the difference that it can only take discrete values since the spherical surface that defines the geometry has a finite size.
The resulting $S(\ell)$ is reported for different temperatures in Fig.~\ref{fig:2}C (see also Fig.~S6).
The behavior of $S(\ell)$ is very similar to that of a standard structure factor in bulk, without all the low-$q$ oscillations due to the underlying spherical geometry. We thus find that $S(\ell)$ is characterized by a main peak for $\ell\sim 18$, followed by smooth oscillations. 
The $T$-dependence of $S(\ell)$ shows that such main peak becomes sharper, shifting to lower and lower $\ell$ as $T$ increases.  Since this calculation is performed on a unit sphere, it is analogous to the rescaled $S(q)$ reported in Fig.~\ref{fig:2}B, thus confirming the shift observed in experiments and in simulations for the full structure factors. To sum up, the shift to low wavevectors of all structure factors clearly indicate that the NPs increase their relative geodesic distance (on the unit sphere) as the microgels undergo the VPT.
This is  due to their mutual electrostatic repulsion that becomes stronger and stronger as they get closer due to the underlying shrinking of the microgel. 
We further assessed the role of electrostatic interactions in determining NPs arrangement and its modifications by simulations at varying the NPs charge (Fig.~S6).
Furthermore, we note that the present data do not show evidence of a crystalline arrangement of NPs, although the main peak in $S(\ell)$ visibly sharpens.
Nevertheless, we set the bases to investigate the dynamics of colloidal phase transitions at the nanoscale that can be achieved by fine tuning of the microgel and NPs characteristics.

\begin{figure*}[t]
\includegraphics[width=0.98\textwidth]{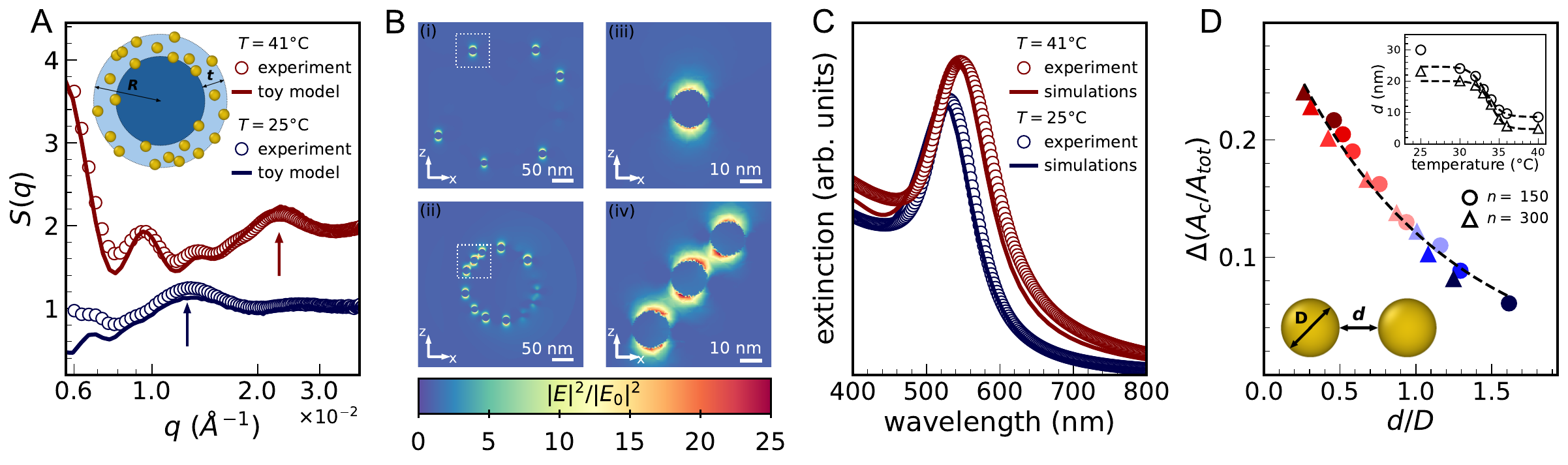}
\caption{\label{fig:3}
(A) Experimental $S(q)$ compared to the predictions of a toy model (sketched in figure), using  $Np=75$, $R=184$ nm, $t=47$ nm, $d_{min}=28$ nm and $\sigma_R=6$ nm for $T=25$°C, and $Np=65$, $R=82$ nm, $t=2$ nm, $d_{min}=8$ nm, $\sigma_R=6$ nm for $T=41$°C; the arrows identify the position of the peaks $q_p$ corresponding to the distance between NPs. (B) Intensity maps of the electric field enhancement $|E|^2/|E_0|^2$, obtained by full-wave numerical simulations for the microgel--NPs complexes, for $\lambda=630$ nm. Cross-sections at the planes $y=39$ nm for $T=25$°C (i) and $y=21$ nm for $T=41$°C (ii). Zoom on a single NP at $T=25$°C (iii) and on a group of NPs at $T=41$°C (iv), highlighting the localization of plasmon coupling.
(C) Simulated extinction spectra at $T=25$°C and $T=41$°C (dashed lines) compared with experimental ones. (D) Degree of coupling $\Delta (A_c/ A_{tot})$ as a function of $d/D$, with $d$ the average surface-to-surface distance and $D$ the NP diameter, for $n=150$ and $n=300$ (same color-coding as in Fig.~2A); the dashed line is a fit to an exponential decay \cite{jain2007}.  Inset: $d$ as a function of $T$. } 
\end{figure*}

To connect the microscopic NP arrangement to the degree of plasmon coupling, we use a simple toy model, sketched in Fig.~\ref{fig:3}A and detailed in SM, which consists in randomly arranging a set of points, mimicking NPs, within a spherical shell, which represents the external corona of the microgels. By using this model, we are able to vary the features of the system arbitrarily, matching the experimental NP/microgel size ratio, which is not possible within the present simulations. In the toy model, we thus vary four different parameters independently: the external radius $R$ and the thickness $t$ of the shell, the number of NPs in the shell $N_p$ and the minimum distance $D+d_{min}$ between them, where $D$ is the NP diameter and $d_{min}$ the minimum surface-to-surface distance. We then calculate $S(q)$ of the NPs within the shell and find that the model clearly allows us to interpret the different features of the measured $S(q)$ and to discern the influence of each parameter on its behavior (Figs.~S7 and S8). Specifically, we confirm that the low-$q$ features arise from the confinement of particles within a spherical shell, and indeed, they change when altering $R$ and $t$.
Instead, the broader band between 1 and 3 $\times 10^{-2}$ \AA$^{-1}$, whose maximum at $q_p$ (highlighted by arrows) increases in intensity and shifts to higher $q$ values with temperature, does not depend on the spherical geometry but rather on $d_{min}$ and $N_p$. This band corresponds to the part of the experimental structure factor that more strongly depends on the NPs mutual interactions.
Fig.~\ref{fig:3}A shows that, with an appropriate choice of parameters, and accounting for the experimental polydispersity $\sigma_R$ as detailed in the SM, it is possible to almost quantitatively reproduce the experimental spectra, in particular matching the position of the various peaks of the experimental $S(q)$. Despite minor deviations in the intensities of the maxima and minima at low $q$, due to the underlying simplicity of the model, we  conclude from this analysis that the more pronounced peaks occurring at intermediate $q$ in experiments are due to the uniforming of the core-corona structure of the microgels as they undergo the VPT. Notably, the model yields meaningful values of the parameters for the description of the experimental results and their trend with temperature. Indeed, we find that the external radius of the shell $R$ shrinks from 184 to 82 nm, in good agreement with $R_H$ measurements reported in Fig.~S3, while the shell thickness $t$ reduces from 47 to 2 nm. This pronounced shrinking of the shell, much greater than the overall microgel deswelling, is due to the incorporation of NPs \cite{brasili2023}, indicating that, across the VPT, the corona of the microgel compacts much more than the core, thus pushing the NPs outward. The further action of mutual electrostatics finally results in the rearrangement of NPs at overall larger relative distances, also favouring an increase ordering, as revealed by $S(\ell)$. 
As further validation of the toy model, we employ the obtained NP configurations to compute the extinction spectra through electromagnetic full-wave simulations, appropriately taking into account the heterogeneous internal microgel structure, as detailed in the SM. 
The resulting maps of electric field enhancement (Fig.~\ref{fig:3}B) clearly show that before the VPT, the NPs on the microgel behave as isolated particles, while afterwards, the electric field is delocalized over adjacent NPs, highlighting the activation of plasmon coupling.
The corresponding spectra, shown in Fig.~\ref{fig:3}C, are in excellent agreement with experimental data, confirming that the proposed model accurately describes both the structural and optical properties of the system.

Based on these results, we can finally connect the experimental $S(q)$ with the optical properties of the complexes as the microgels undergo the VPT. Starting from the position $q_p$ of the first peak of the effective  structure factor of the NPs, which is not related to the spherical confinement but reflects the local interparticle arrangement (see Fig.~\ref{fig:3}A), we determine the average surface-to-surface distance $d=2\pi/q_p - D$ between nearest neighbors, with $D=18.6$ nm the NP diameter (Fig.~S2), reported as a function of $T$ in the inset of Fig.~\ref{fig:3}D. We then report the degree of coupling, obtained from the extinction spectrum at the same $T$, as a function of the ratio $d/D$ to quantify the plasmons interaction strength \cite{jain2007} (Fig.~\ref{fig:3}D). Since NPs approach each other more and more upon shrinking of the microgels above the VPT, they are able to explore distances of just a few nanometers between their  surfaces, and thus to give rise to plasmon coupling.
Notably, we find that the data for the two studied values of $n$ follow the same trend, compatible with an exponential decay pattern~\cite{jain2007} with a characteristic length $\sim$ 20 nm. 
Altogether, these results shed light for the first time on the direct relationship between the coupling of plasmonic NPs adsorbed onto thermoresponsive microgels and their average inter-particle distance. This enables the optical measurement of the NP-NP nanoscale distances on the spherical surface as a function of temperature, thus serving as a plasmonic ``nano-ruler'', similar to those based on DNA technology~\cite{sonnichsen2005}. Our study thus opens the possibility to rationally design of thermoresponsive plasmonic systems with optical properties tailored to specific needs. For instance, by adjusting the chemical composition of microgels and NP adsorption, this system can serve as a reactor with thermally tunable catalytic activity \cite{liu2015electrostatic,chang2023synergistic}, with the additional advantage of a real-time monitoring of NPs distance.

In summary, the present work reports a detailed structural analysis of the spatial organization of plasmonic NPs electrostatically adsorbed on thermoresponsive microgels and of its evolution across the VPT. By combining SAXS and numerical simulations, supported by a simple toy model, we reveal a clear tendency of the NPs to maximize their relative distance on the spherical surface of the microgel due to the underlying electrostatic repulsion. 
Importantly, we also provide a link between the structure of the microgel--NPs complexes and their optical properties, confirmed by electromagnetic full-wave simulations, establishing for the first time the relationship between the degree of plasmon coupling in the extinction spectra and the surface-to-surface distance between neighbour NPs.
In addition, our study provides a paradigm shift in the use of thermoresponsive microgels, which serve as a tunable substrate for the colloidal assembly in curved two-dimensional geometry, opening the way to tackle experimentally this important physical problem at the nanoscale.

We thank A.L. Božič for valuable discussions. We acknowledge SOLEIL for providing synchrotron radiation facilities under proposal n. 20191601 at the SWING beamline and the CINECA award under the ISCRA initiative, for the availability of high-performance computing resources and support.
F.Br., F.B., E.Z. and S.S. acknowledge financial support from INAIL, project MicroMet (BRiC 2022, ID 16). E.Z. and S.S. also acknowledge support from ERC POC project MICROSENS (grant agreement no.101157420). 
A.C. and C.R. acknowledge financial support from the European Union - NextGenerationEU under the Italian Ministry of University and Research (MUR) National Innovation Ecosystem grant ECS00000041 - VITALITY - CUP E13C22001060006.
D.T. and E.C. acknowledge financial support from the Agence Nationale de la Recherche (Grant ANR-20-CE06-0030-01; THELECTRA).

\newpage
\begin{center}
\section*{Supplemental Material}
\end{center}
\setcounter{equation}{0}
\setcounter{figure}{0}
\setcounter{table}{0}
\setcounter{section}{0}
\renewcommand{\theequation}{S\arabic{equation}}
\renewcommand{\thefigure}{S\arabic{figure}}
\renewcommand{\thetable}{S\arabic{table}}
\renewcommand{\thepage}{S\arabic{page}}
\renewcommand{\thesection}{S\arabic{section}}

\section{Experimental methods}
\paragraph*{Microgel synthesis.}
Our experiments are performed on cationic microgels with fraction of crosslinker monomer $c=0.05$ and of initiator $f/2=0.01$, synthesized by the surfactant-free radical polymerization previously detailed \cite{truzzolillo2018, brasili2023}.
We dissolve 1.25 g of NIPAM monomers (Sigma-Aldrich, $\text{MW}=113.16$ Da) and the crosslinker N,N'-methylene-bis-acrylamide (BIS, Sigma-Aldrich, $\text{MW}=154.17$ Da) in 148 mL of deionized water. Separately, the ionic initiator 2,2'-Azobis(2-methylpropionamidine) dihydrochloride (AIBA, Sigma-Aldrich, $\text{MW}=271.19$ Da) is dissolved in 2 mL of water.
The solution containing NIPAM and BIS is bubbled with argon for 30 minutes and, after heating up to 70 °C, the initiator solution is added. 
At 70 °C, AIBA undergoes homolytic cleavage forming two radicals, each including one amine group. Each radical reacts with a NIPAM monomer and produces a new radical, giving rise to the polymerization reaction. Therefore, after starting the reaction, amines of AIBA initiator remain attached to the backbone of the microgels, providing them positive charge due to protonation. 
After 6--hour reaction, the obtained dispersion is cooled down to room temperature and filtered through glass wool. 
To prevent bacterial growth, \ce{NaN3} (Sigma-Aldrich, $\text{MW}=65.01$ Da) is added to the concentration of 2 mM.
The hydrodynamic radius of the microgels, measured by dynamic light scattering (DLS) at 25 °C, is $R_H = 286$ nm. 
The final number density of the microgels in the dispersion is evaluated to $n_{mg} = 1.63 \times 10^{12}$ mL$^{-1}$ by viscosimetry measurements as described in Refs.~\citenum{brasili2023} and \citenum{truzzolillo2015}.

\paragraph*{Preparation of microgel--nanoparticles samples.}
We use spherical gold nanoparticles (NPs, Ted Pella) with nominal diameter $D=20$ nm and number density $n_{NP}=7.0\times10^{11}$ mL$^{-1}$. A stabilizing citrate capping provides NPs with a negative charge, previously evaluated to $q_{NP}=-35\,e$ \cite{brasili2023}. The very low concentration guarantees that possible effects on the ionic strength of the final samples due to ions release from the NPs surface are negligible.
To prepare samples, we dilute separately the microgel dispersion 250 times in 0.4 mM \ce{NaN3} and the NPs one in water to obtain the desired number ratio $n = n_{NP}/n_{mg}$. We then mix the two components and gently agitate the solution by hand.
The concentration of \ce{NaN3} in the final samples is 0.2 mM, low enough to exclude any effect of the ionic strength on the microgel swelling. At the studied conditions, the Debye length is estimated to be $\sim$ 20 nm.

Since we are interested in studying the plasmon coupling, that takes place for surface-to-surface distances between NPs of few nanometers, we choose high $n$ values to achieve sufficiently small distances and to avoid inducing aggregation of the microgel--NPs complexes. Preliminary DLS and extinction spectroscopy experiments show that these conditions are matched for $n\geq 100$. Moreover, since for high $n$ only part of the NPs in the sample is actually incorporated within the microgel network due to the electrostatic repulsion \cite{brasili2023}, we need to ensure that the portion of non-adsorbed NPs is not predominant to avoid compromising the outcomes of the experiments. A previous study of NPs adsorption on microgels with $f=0.032$ \cite{brasili2023} shows that this requirement is well-fulfilled for $n \leq 300$. 
Even though the microgels in this study are slightly less charged, we adhered to these limits, selecting $n=150$ and $n=300$. This choice is further supported by the fact that we were able to acquire extinction and scattering intensity spectra with a good signal, suitable for the analyses of this study.

\paragraph*{Extinction spectroscopy.}
Extinction spectra in the UV–Visible-NIR spectral range were acquired by including the samples in a 1 mm quartz cuvette and using a v-570 double ray spectrophotometer (Jasco, Tokyo, Japan), equipped with a Peltier thermostatted holder EHC-505 (Jasco). The instrument has a spectral resolution of 0.1 nm in the UV–Visible range and 0.5 nm in the NIR range. For measurements at varying temperature, samples were kept thermalising 5 minutes after every temperature change before acquisition. 
All the spectra reported in the figures are normalized to the extinction at 400 nm, that corresponds to the onset of the gold interband transitions and is unaffected by particle size, shape, and environment, contrarily to the LSPR \cite{amendola2017}.

\paragraph*{Small Angle X-ray Scattering.}
For Small Angle X-ray Scattering (SAXS) experiments, samples were filled in capillaries (1.5 mm diameter) and placed at the sample-to-detector distance of 3 m. 
The exposure time for acquisitions was set to 1 s and 14 scattering patterns were acquired for each sample. Scattering patterns were recorded  at 12 keV using a two-dimensional EigerX 4-M detector (Dectris, Baden, Switzerland). This allows measurements in the range of $q$-vector between 0.002 and 0.38 \AA$^{-1}$, where $q$ is defined as $q=(4\pi/\lambda)\sin\theta$, $2\theta$ is the scattering angle, and $\lambda$ is the wavelength of the radiation. Scattering patterns of an empty capillary and of a capillary filled with water were recorded for normalization of the intensity to absolute units and background subtraction, respectively.
Experiments were conducted at selected temperatures between 25 and 41 °C employing a Huber Ministat 125 thermostat. After each temperature change, samples were left thermalizing for 5 minutes before measurements. The processing and averaging of the scattering patterns were performed by the software Foxtrot (SOLEIL software group and SWING beamline). When averaging, any scattering curve not perfectly superimposed with the overall set acquired, due to possible residual equilibration or other experimental perturbations, was discarded.

For a collection of particles, the scattered intensity $I(q)$ can be expressed in terms of the form factor $P(q)$ of single particles and of the structure factor $S(q)$ of the system as $I(q)=nv^2\Delta\rho^2P(q)S(q)$, where $n$ and $v$ are the number density and the volume of the scattering particles, and $\Delta\rho$ is the contrast in electron density $\rho$ between particles and solvent.
$S(q)$ is the interference introduced by interparticle correlations and can be expressed in terms of the Fourier transform of the pair correlation function $g(r)$ as:
\begin{equation}\label{eq:structure_factor}
S(q) = 1 + \rho \int_V g(r) e^{-i\vec{q}\cdot\vec{r}} d \vec{r} \quad .
\end{equation}
Since for a dilute system of non-interacting scatterers $S(q)=1$, we directly measured the form factor of gold NPs on the stock solution \cite{capocefalo2022}. 
In microgel--NPs samples, given the extremely higher contrast $\Delta\rho$ of gold than that of the polymer chains, it is reasonable to assume that the measured scattered intensities, with the chosen acquisition times, originates solely from the NPs present in the sample. 
We verified this assumption by measuring a microgel sample without NPs in the same experimental conditions, resulting in almost null scattered intensity.
We therefore could simply derive the structure factor of NPs by dividing the scattered intensity measured on each sample by the form factor of NPs.

\paragraph*{Dynamic Light Scattering.}
Size distributions in terms of hydrodynamic radius $R_H$ were measured by Dynamic Light Scattering (DLS), employing a NanoZetaSizer apparatus (Malvern Instruments LTD) equipped with a \ce{He-Ne} laser (5 mW power, 633 nm wavelength) that collects light in quasi-backscattering geometry, at an angle of 173$^{\circ}$. Decay times, extrapolated from the acquired intensity autocorrelation functions, are used to determine the distribution of diffusion coefficients $D_T$ of the particles. Diffusion coefficients are then converted to intensity-weighted distributions of $R_H$ using the Stokes-Einstein relationship $R_H = k_BT/6\pi\eta D_T$, where $k_BT$ is the thermal energy and $\eta$ the water viscosity. 
Temperature trends are measured using ascending ramps between 25 $^{\circ}$C and 41 $^{\circ}$C. After each temperature variation, the samples were kept thermalizing for 5 minutes before performing the measure.
Each value of $R_H$ reported in the work is the average of a distribution obtained by at least 50 measurements. The associated error is the corresponding standard deviation.

\section{Additional Experiments}

\begin{figure}[ht]
\includegraphics[width=0.98\textwidth]{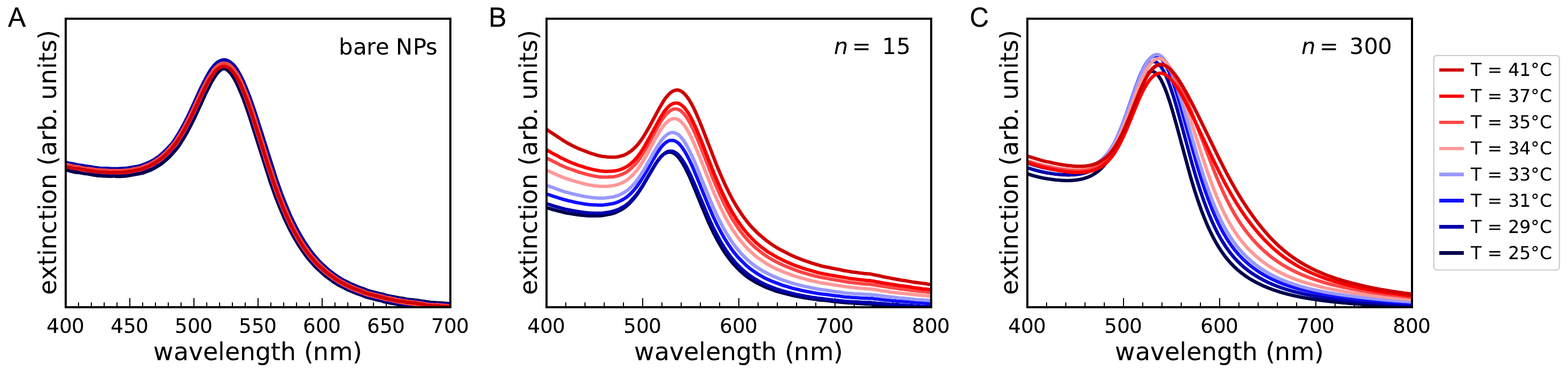}
\caption{ \label{fig:figS1} Extinction spectra of the NPs stock dispersion (A) and of the microgel--NPs complexes as a function of temperature for $n=15$ (B) and $n=300$ (C).}
\end{figure}

\paragraph*{Optical analysis: extinction spectra and plasmon coupling.}
We report in Fig.~\ref{fig:figS1} the extinction spectra as a function of temperature between 25°C and 41°C of the NPs stock dispersion and of microgel--NPs complexes with number ratios $n=15$ and $n=300$.
Due to the high dilution and colloidal stability of the stock sample, no plasmon coupling is observed for NPs alone (Fig.~S1A) in the overall range of temperatures, as the extinction spectrum remains unaltered.
For $n=15$, the adsorbed NPs are few enough to remain far apart even above the VPT. Since in these conditions no plasmon coupling occurs, this sample is a reference to analyze the effects on the LSPR caused by water expulsion from the polymer network and consequent chain densification across the VPT. Hence, the observed shift of the LSPR, up to 8 nm, can be attributed solely to the variation of refractive index at the interface of the NPs with increasing temperature.
At larger $n$ values, the maximum shift is higher, reaching 10 nm for $n=300$ and 16 nm for $n=150$, pointing to a contribution from plasmon coupling.
However, even if the shift of the LSPR gives a good description of the optical changes across the VPT and can be used to connect the optical properties of microgel--NPs complexes and the structure of the polymer network \cite{zygadlo2024}, it does not reveal any information about broadening of the peak and onset of coupled plasmon modes at higher wavelengths, that are the main features of interest for many applications, such as colorimetric detection or surface-enhanced spectroscopies \cite{aili2011}. To account for these changes, clearly distinguishable in the spectra, we therefore choose a more robust method to quantify the plasmon coupling. Analogously to refs.~\citenum{aili2011} and \citenum{chowdhury2004}, we identify the spectral region of coupled modes in the wavelength range from 570 to 800 nm (inset of Fig.~1A), where only low-energy modes resulting from near-field interactions contribute to the extinction, and calculate their spectral weight.
We then define the coupling degree as 
\begin{equation}\label{eq:delta_area}
\Delta\frac{A_C}{A_{tot}}=\frac{A_C(T)}{A_{tot}(T)}-\frac{A_C^{(NP)}}{A_{tot}^{(NP)}}
\end{equation}
where $A_{tot}(T)$ and $A_C(T)$ are the areas underlying the spectrum acquired at temperature $T$ on the microgel--NPs samples, in the overall spectral range and in the region of coupled modes, respectively; $A_{tot}^{(NP)}$ and $A_C^{(NP)}$ are the corresponding areas computed from the reference spectrum of NPs stock dispersion, which due to its high dilution and colloidal stability ensures the absence of plasmon coupling.

\begin{figure}[ht]
\centering
\includegraphics[width=0.8\textwidth]{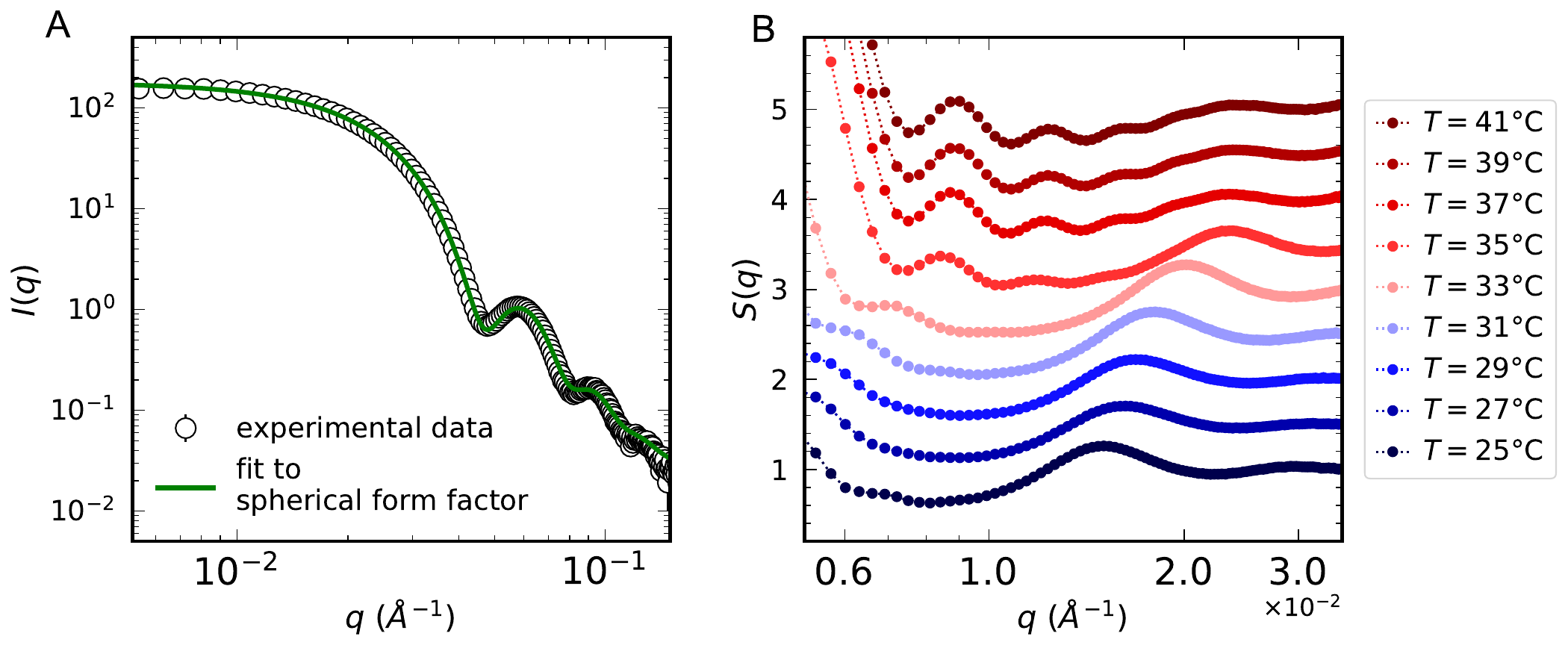}
\caption{(A) Experimental form factor of gold NPs at $T=25$ °C and best fit to a sphere model. (B) Structure factors of NPs as a function of temperature for the microgel--NPs sample with $n=300$.}
\label{fig:figS2}
\end{figure}

\paragraph*{SAXS analysis: form factor of gold NPs and structure factors of microgel--NPs.}
The form factor of gold NPs is directly measured by SAXS on dilute samples, with number density $n_{NP}=7.0\times10^{11}$ mL$^{-1}$. The acquired scattering curve is reported in Fig.~\ref{fig:figS2}A.
We fitted the curve to a spherical form factor model \cite{guinier1955} with log-normal particle size distribution, to evaluate the radius of NPs to 9.3 nm with 10\% polydispersity.
The good overlap between the two curves also ensures the validity of the assumption that there is no interaction between the dispersed NPs, thus the scattering curve represents their form factor.

In Fig.~\ref{fig:figS2}B, we report the structure factors of NPs measured on microgel--NPs samples with with number ratio $n=300$ as a function of temperature between 25 °C and 41 °C.

\begin{figure}[ht]
\centering
\includegraphics[width=0.55\textwidth]{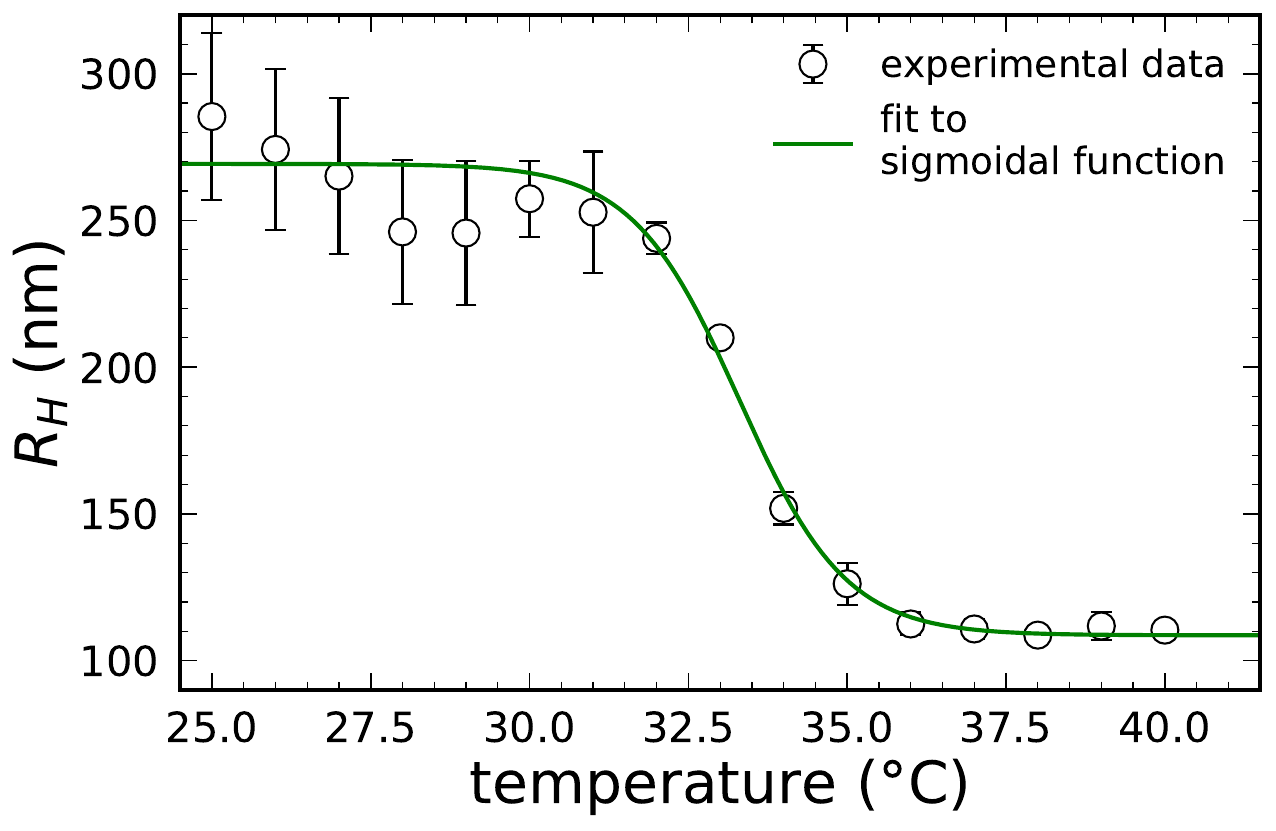}
\caption{Swelling curve of the microgel and best fit to a sigmoidal-like function.}
\label{fig:figS3}
\end{figure}

\newpage
\paragraph*{Swelling curves of microgels.}
The experimental swelling curve of the microgels, obtained plotting $R_H$ as a function of temperature is reported in Fig.~\ref{fig:figS3}. Based on it, we extrapolated the transition temperature $T_C=33.3\pm0.3$ °C, by fitting data to the sigmoidal-like function
\begin{equation} \label{eq:sigmoid}
y=y_{\infty}+\frac{y_{0}-y_{\infty}}{1+e^{\frac{x-T_C}{\Delta T}}} \quad .
\end{equation}

\newpage
\section{Numerical simulations}

\paragraph*{Modelling microgel--NPs.}
We use coarse-grained microgels consisting of fully-bonded, disordered polymer networks of $N$ spherical beads with diameter $\sigma$ and mass $m$, which set the length and mass units. Nanoparticles are also spherical beads of diameter $D$ and mass $m$.

Microgels of $N=112k$ or $N=14k$ monomers are prepared with the protocol previously reported in Refs.~\citenum{gnan2017} and \citenum{ninarello2019}, which was found to reproduce very well the experimental structure of the particles. After assembly,  monomers interact via the bead-spring model, established by Grest and Kremer~\cite{grest1986}, which amounts to the sum of a steric repulsion for all beads plus a bond term for the connected ones.  The first term is modelled by the Weeks--Chandler--Anderson (WCA) potential:
\begin{equation}
\label{eq:wca}
V_{\text{WCA}}(r)  =  
\begin{cases}
4\varepsilon\left[\left(\frac{\sigma}{r}\right)^{12}-\left(\frac{\sigma}{r}\right)^6\right]+\varepsilon & \qquad \text{if} \quad r \le 2^{1/6}\sigma  \\
0 & \qquad \text{if} \quad r > 2^{1/6}\sigma
\end{cases}
\end{equation}
where $r$ is the center-to-center distance between a given pair of interacting particles and $\varepsilon$ sets the energy scale.
In addition, bonded monomers interact via the Finitely Extensible Nonlinear Elastic (FENE) potential \cite{del2019,kremer1990}:
\begin{equation}
\label{eq:fene}
V_{\text{FENE}}(r)  = 
-\epsilon k_F {R_F}^2\ln \left[1-{\left(\frac{r}{R_F\sigma } \right)}^2 \right]\; , \quad r < R_F \sigma\end{equation}
with $R_F=1.5$ and $k_F=15$. Monomers are linked via the FENE potential to two neighbours representing segments of NIPAM chains, whereas crosslinkers have a fourfold valence. As for experiments, we use the molar fraction of crosslinker $c = 0.05$ and bonds, once assembled, cannot break during the course of a simulation. 

To model the VPT of microgels, we introduce an effective solvophobic potential $V_\alpha$, acting only between divalent monomers, which implicitly accounts for monomer–solvent interactions \cite{soddemann2001}:
\begin{equation}
\label{eq:Va}
V_\alpha(r)  =  
\begin{cases}
-\varepsilon\,\alpha & \qquad \text{if} \quad r \le 2^{\frac{1}{6}}\sigma  \\
\frac{1}{2}\,\alpha\,\varepsilon\left[cos (\gamma_0(r\sigma)^2+\beta_0)-1 \right] & \qquad \text{if} \quad 2^{\frac{1}{6}}\sigma  < r \le R_F\sigma \\
0 & \qquad \text{if} \quad r> R_F\sigma
\end{cases}
\end{equation}
where $\gamma_0=\pi(2.25-2^{1/3})^{-1}$ and $\beta_0=2\pi-2.25\gamma_0$.
This term represents an effective attraction, modulated by the solvophobic parameter $\alpha$, which plays the role of an effective temperature. Therefore, $\alpha=0$ represents good solvent conditions (at low temperature, below the transition), while as $\alpha$ rises, the attraction between monomers grows, leading to aggregation and microgel shrinkage; the overall behavior echoing the worsening of the monomer affinity to the solvent when temperature is risen \cite{gnan2017,moreno2018}.

To mimic the ionic groups of AIBA monomers, we provide a fraction $f$ of the microgel beads with a positive charge. Similar to experiments, we simulate microgels with surface charge distribution, where we assign charged beads randomly, but only in the exterior corona of the microgel, i.e. where the distance from the microgel centre of mass is higher than $R_g$. To ensure the overall electro-neutrality, for each charged monomer we also insert an oppositely charged counterion, whose diameter is set to $\sigma_c=0.1\,\sigma$~\cite{del2019}. Counterions interact among each other and with microgel beads through the WCA potential. Additionally, all charged particles interact with the Coulomb potential:
\begin{equation}
V_{\rm coul}(r_{ij})=\frac{q_i q_j \sigma}{e^{*2}\, r_{ij}}\,\varepsilon\; ,
\end{equation}
where $q_i$ and $q_j$ are the charges of the interacting beads ($+\,e^*$ for charged monomers of the microgel and $-\,e^*$ for counterions), being $e^*=\sqrt{4\pi\epsilon_0\epsilon_r\sigma\,\varepsilon}$ the reduced charge unit and $\epsilon_0$, $\epsilon_r$  the vacuum and relative dielectric constants, respectively. The particle-particle-particle-mesh method \cite{deserno1998} is adopted to appropriately account for the long-range nature of the Coulomb interactions.
Following our previous works~\cite{del2020,del2021}, charged monomers on the microgel do not interact with the solvophobic potential, even when temperature increases, to ensure their maintained hydrophilic character in the whole investigated temperature range.

Finally, NPs are represented as single beads with negative charge $q=-\,35\,e^*$ of diameter $D$. To maintain the same proportion between NPs and microgel sizes as in the experiments, we use two values of the diameter, $D=4\,\sigma$ for microgels with $N=112k$ and $D=2\,\sigma$ for $N=14k$. 
Similarly to when assigning charge to AIBA monomers, an appropriate number of positive counterions ($q=+e^*$, $\sigma_c=0.1\,\sigma$) are added to preserve for all studied conditions the overall neutrality of the system.
NPs, as all charged beads, interact among each other and with all other beads through the WCA and Coulomb potentials. For all mixed interactions,  the interaction lengths are given by the mixed diameters $\sigma_{ab}=(\sigma_a+\sigma_b)/2$, with $a,b$ indicating all different species involved (NPs, monomers either charged or uncharged and all counterions).

NVT simulations are performed with the LAMMPS package~\cite{plimpton1995} at the temperature fixed by $k_BT=\varepsilon$ in a cubic box with side $L$ and periodic boundary conditions.
We choose $L = 600\,\sigma$ for $N=112000$ and $L = 
300\,\sigma$ for $N=14000$.
The equations of motion are integrated with a time-step $\Delta t = 0.002\,\tau$, where $\tau = \sqrt{m\sigma^2/\varepsilon}$ is the reduced time unit. We use the Nos\'e-Hoover thermostat in the constant NVT ensemble for equilibration ($1000\,\tau$) and the Velocity-Verlet algorithm in the constant-energy ensemble for the production runs ($20000\,\tau$). The latter are used to extract the equilibrium averages of the observables of interest.

The size of microgel--NPs complexes is characterized in terms of hydrodynamic radius $R_H$, computed from simulations using the ZENO software \cite{zeno}. We include in the calculation all the monomers of the microgel and all NPs bound to charged monomers, identified as those with distance from the closest charged monomer is lower than $1.25\,(D+\sigma)$.

The structure factors of adsorbed NPs are calculated at each wavenumber $q$ as:
\begin{equation}\label{eq:Sq}
S(q)= \frac{1}{N_{\text{ads}}}\sum_{i=1}^{N_{\text{ads}}}\sum_{j=1}^{N_{\text{ads}}} e^{-i\vec{q}\cdot\vec{r}_{ij}} \quad,
\end{equation} 
where $\vec{r}_{ij}$ is the distance between the $i$-th and $j$-th NPs, and the sum is performed over all the $N_{\text{ads}}$ NPs adsorbed to the microgel, identified as those whose distance from the microgel center of mass is lower than $R_H+D/2$.

The spherical structure factors $S(\ell)$ are calculated using the geodesic distances between the NPs, \textit{i.e.}~the angle determined by their positions on the unit spherical surface centered at the microgel's center of mass. For an $(i,j)$  pair of NPs, located at $\vec{r}_i$ and  $\vec{r}_j$, the geodesic distance $\gamma_{ij}$ is given by:
\begin{equation}
    \gamma_{ij} = 2\arcsin(\frac{1}{2} |\hat{r}_i-\hat{r}_j| ) \quad ,
\end{equation}
where $\hat{r}_{i/j}$ is the versor of $\vec{r}_{i/j}$. $S(\ell)$ is then defined as \cite{bozic2019}: 
\begin{equation}\label{eq:Sl}
S(\ell) = 1 + \frac{2}{n} \sum_{(i,j)} P_\ell(\cos\gamma_{ij}) \quad ,
\end{equation}
where the sum is performed over all  $(i,j)$ pairs and $P(\ell)$ is the Legendre polynomial of degree $\ell$. 
Here, $\ell$ plays an equivalent role to the wave vector $q$ in ordinary space, with the difference that it can only take discrete values since the spherical surface that defines the geometry has a finite size.

\begin{figure}[ht]
\centering
\includegraphics[width=0.55\textwidth]{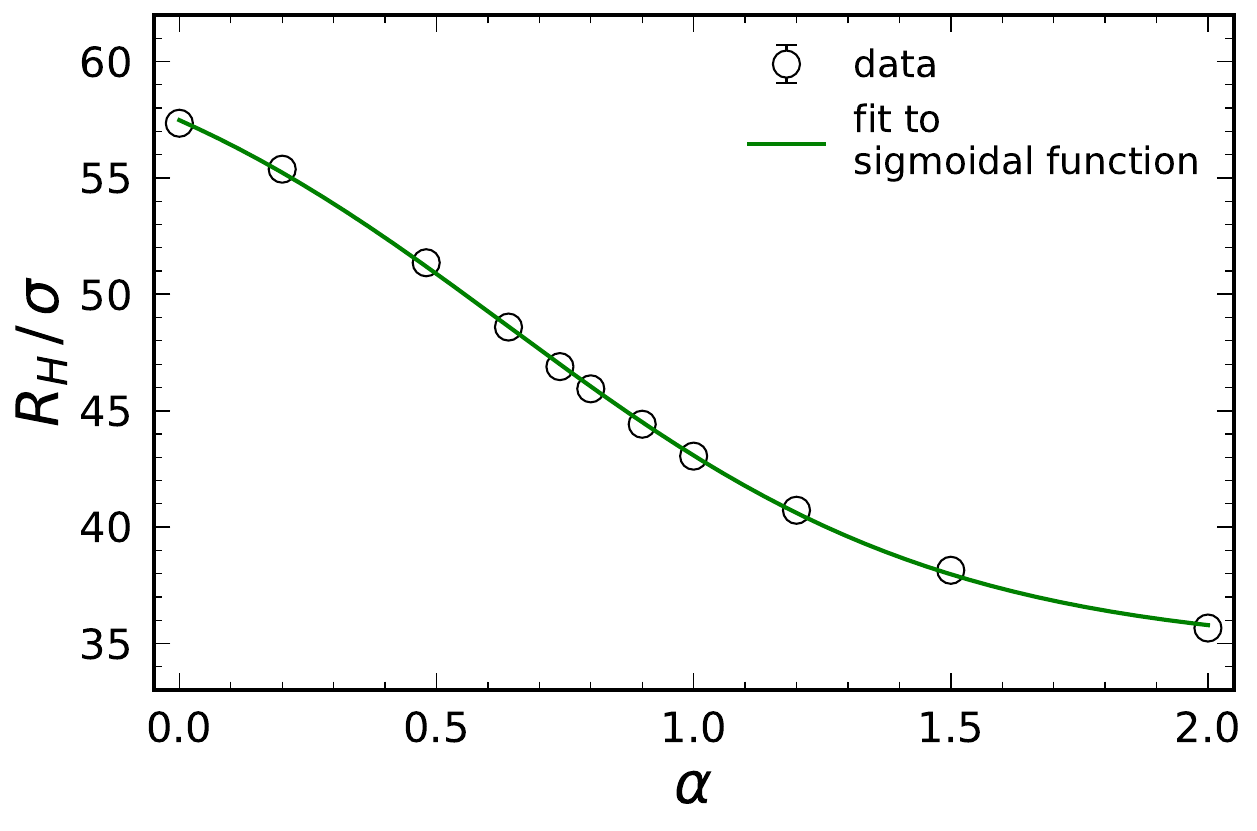}
\caption{Swelling curve, $R_H$ as a function of $\alpha$, of the microgel with $N=112k$ and $f=0.02$, and best fit to a sigmoidal-like function.}
\label{fig:figS4}
\end{figure}

\paragraph*{Swelling curve.}
The swelling curve of the simulated microgel for $N=112k$ and $f=0.02$, as obtained by plotting $R_H$ as a function of $\alpha$ is reported in Fig.~\ref{fig:figS4}. Based on it, we extrapolate the equivalent transition temperature $\alpha=0.64\pm0.02$, fitting the  data to the sigmoidal-like function of Eq.~\ref{eq:sigmoid}.

\begin{figure}[hb]
\centering
\includegraphics[width=0.98\textwidth]{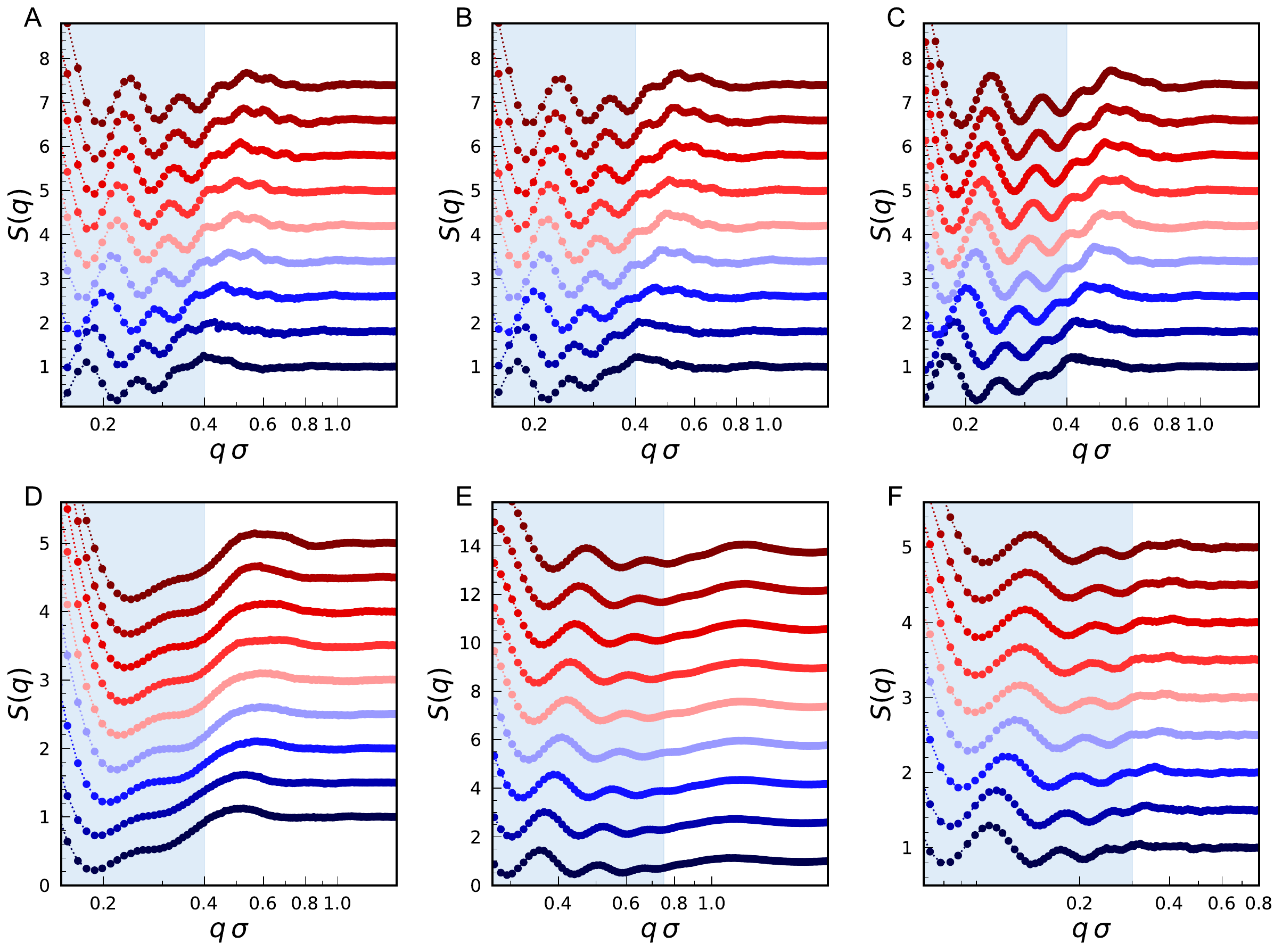}
\caption{Structure factors at varying $\alpha$ for different microgel--NPs systems: microgel with $N=112k$ and $f=0.02$, for $n=80$ (A), $n=100$ (B) and $n=300$ (C), and microgels with $N=14k$ and $f=0.02$ (D), $N=14k$ and $f=0.16$ (E), $N=112k$ and $f=0.01$ (F) at fixed $n=150$. The color-coding is the same as in Fig.~1.}
\label{fig:figS5}
\end{figure}

\newpage
\paragraph*{Structure factors of NPs adsorbed to microgels.}
The NPs structure factors for the microgels with $N=112k$ and $f=0.02$ at different NP/microgel number ratios and for different microgel compositions ($N=14k$ and $f=0.02$, $N=14k$ and $f=0.16$, $N=112k$ and $f=0.01$) at fixed $n=150$ are reported as a function of $\alpha$ in Fig.~\ref{fig:figS5}.
In all the cases considered, the high oscillations at low $q$ are followed by smaller peaks at intermediate $q$ (highlighted in light blue). These peaks are present for all the effective temperature, in contrast to experiments.

\begin{figure}[hb]
\centering
\includegraphics[width=0.95\textwidth]{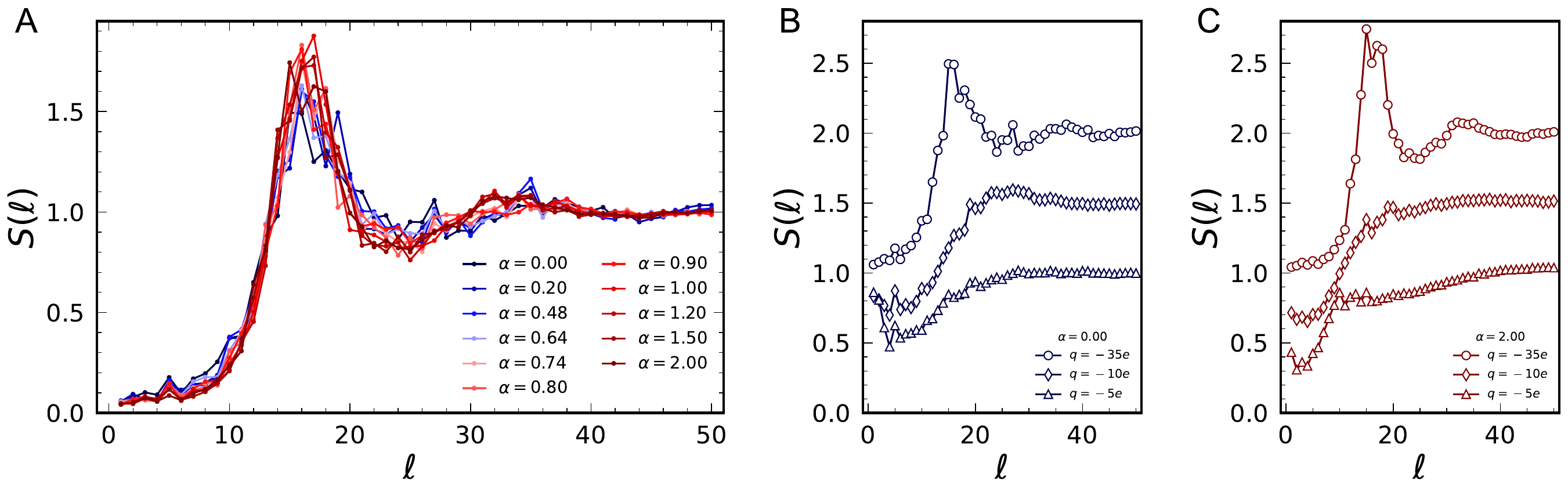}
\caption{Spherical structure factors at varying $\alpha$ for the microgel--NPs complex with $N=112k$, $f=0.02$ and $n=150$ (A). Spherical structure factors for the same microgel at varying the NP charge $q$, below (B, $\alpha=0.00$) and above (C, $\alpha=2.00$) the VPT of the microgel.}
\label{fig:figS6}
\end{figure}

\newpage
\paragraph*{Spherical structure factors of NPs adsorbed to microgels.}
The spherical structure factors of the adsorbed NPs for the system with $N=112k$, $f=0.02$ and $n=150$, computed according to Eq.~\ref{eq:Sl}, are reported as a function of $\alpha$ in Fig.~\ref{fig:figS6}A. 
We find that $S(\ell)$ is characterized by a main peak for $\ell\sim 16$, followed by smooth oscillations. Additionally, it approaches a low value as $\ell \rightarrow 0$. 
The $T$-dependence of $S(\ell)$ shows that the main peak becomes sharper, shifting to lower $\ell$ as $T$ increases. The same variations are more pronounced in the case of the microgel with $N=112k$ and $f=0.16$ reported in Fig.~2C.
To sum up, the shift to low wavevectors of all structure factors clearly indicate that the NPs increase their relative (geodesic) distance as the microgels undergo the VPT.

In order to better assess the role of electrostatic interactions, we performed simulations on the same microgel interacting with NPs with different charges. We considered the two additional values of the NP charge, $q=-\,5\,e^*$ and $q=-\,10\,e^*$. The plots of $S(\ell)$ for the three values of $q$ are reported in Fig.~\ref{fig:figS6}B and C, for $\alpha=0.00$ and $\alpha=2.00$, respectively.
Noteworthy, the main peak of the curve, that becomes narrower and more pronounced as $\alpha$ increases above the VPT, is observable only for $q=-\,35\,e^*$, when the number of adsorbed NPs is higher and the electrostatic repulsion between them is stronger. This highlights the important role of electrostatic interactions in determining the spatial arrangement of the adsorbed NPs and its modification across the VPT.

\newpage
\section{Toy model}
\paragraph*{Description of the model}
The toy model, sketched in Fig.~3A, consists in a set of particles randomly positioned within a spherical shell; the particles mimic the NPs and the shell they are restrained represents the external corona of the microgel.
The system is built from four parameters: the external radius $R$ of the shell, its thickness $t$, the target number of particles to be placed within the shell and the minimum surface-to-surface  distance $d_{min}$ between any two particles. 
The procedure to place the $i$th particle inside the shell is the following: first, (\textit{i}) a random position within the shell is generated; next, (\textit{ii}) the distance of this newly created position to all $i-1$ previously inserted particles is calculated; (\textit{iii}) if the distance is larger than $D+d_{min}$ for all present particles (here $D=18.6$ nm is the NP diameter, obtained by the fit of Fig.~S2), the $i$th particle is assigned to this position. 
The procedure is repeated in case of failing to allocate a new particle for a maximum number of $50k$ attempts for each particle. The limiting maximum number of attempts implies that in systems with high surface densities the number $N_p$ of particles actually placed in the shell may be lower than the target one.

For each set of the four parameters, we prepare 100 distinct configurations and compute the average of the corresponding structure factors, according to Eq.~\ref{eq:Sq}.
To account for size polydispersity of experimental microgels, we consider a Gaussian distribution for the shell radius, defined by the mean value $R_0$ and standard deviation $\sigma_R$. We then extract from the distribution a different value of $R$ for each of the 100 configurations; to keep constant the geometrical proportions, the shell thickness is also corrected by a factor $R/R_0$.
In this way the features of the average structure factor result more smoothed and therefore overlap better with experimental data.

\begin{figure}[ht]
\centering
\includegraphics[width=0.96\textwidth]{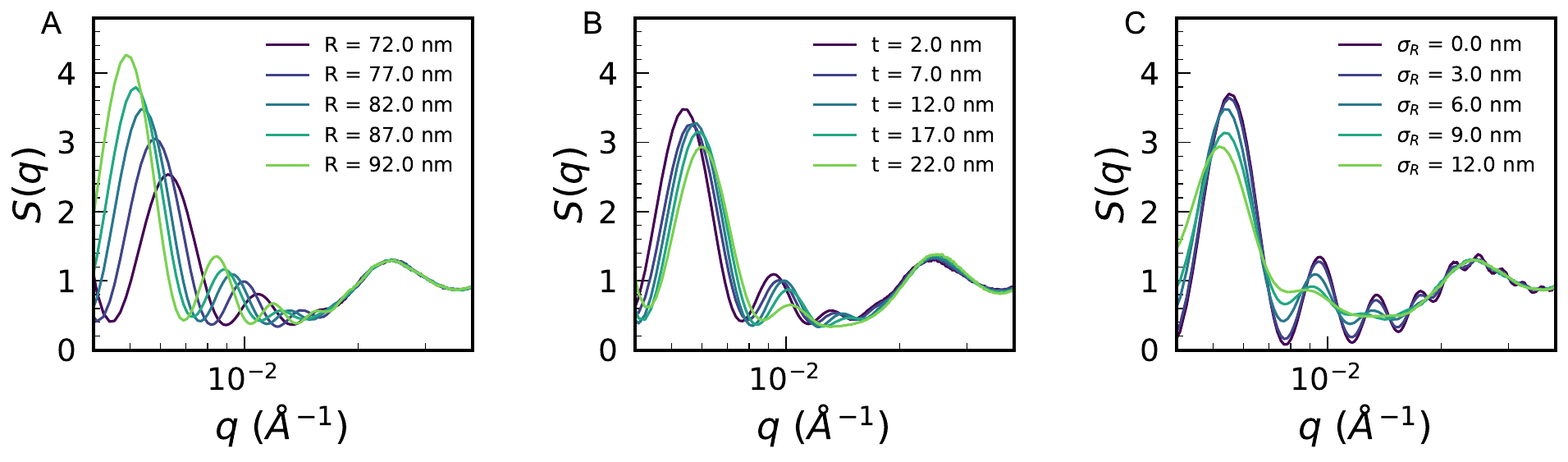}
\caption{Structure factors of the toy model, consisting in particles with random positions within a spherical shell, at varying independently each geometrical parameter of the shell $R$ (A), $t$ (B) and $\sigma_R$ (C). The following values are used for the parameters that are kept constant: $R=82$ nm, $t=2$ nm, $\sigma=6$ nm, $N_p=150$ and $d_{min}=$ 8 nm. \label{fig:figS7} \vspace{2cm}}
\includegraphics[width=0.64\textwidth]{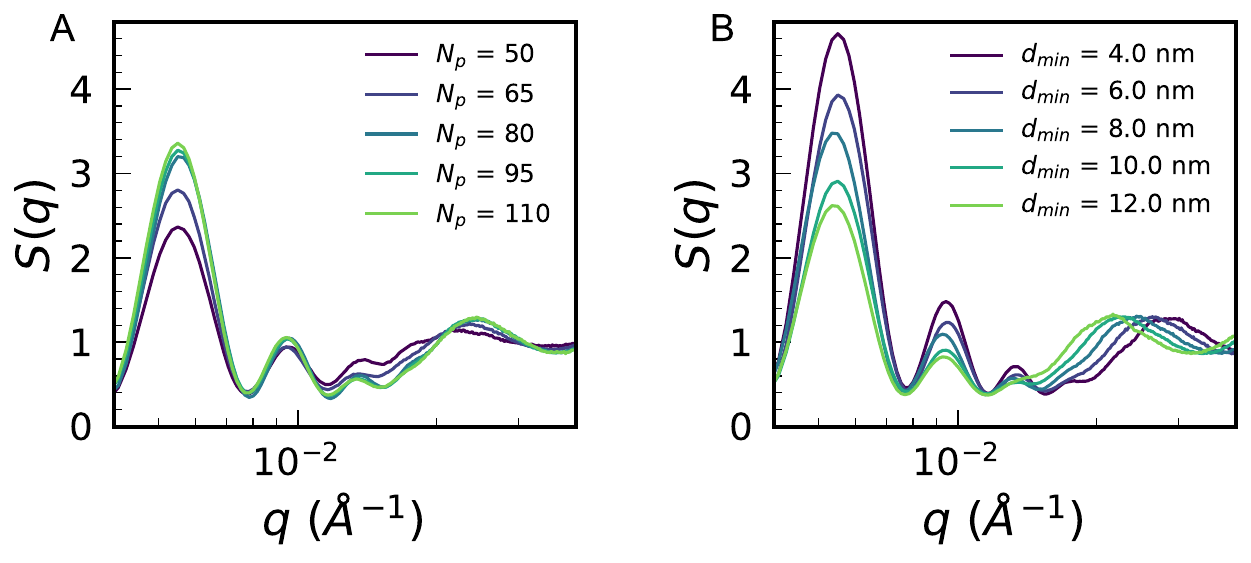}
\caption{Structure factors of the toy model, consisting in particles with random positions within a spherical shell, at varying independently each parameter of the NPs, $N_p$ (A) and $d_{min}$ (B). The following values are used for the parameters that are kept constant: $R=82$ nm, $t=2$ nm, $\sigma=6$ nm, $N_p=150$ and $d_{min}=$ 8 nm. \label{fig:figS8}}
\end{figure}

We used the toy model to achieve a full comprehension of the structure factors of the NPs adsorbed to microgel. Specifically, we varied each of the model parameters independently to understand how it affects the features of the $S(q)$. The trends at varying the shell parameters $R_0$, $t$ and $\sigma_R$ are shown in Fig.~\ref{fig:figS7}, those at varying the NPs parameters $N_p$ and $d_{min}$ are shown in Fig.~\ref{fig:figS8}.

\clearpage
\section{Full-wave simulations}
\begin{figure}[ht]
\centering
\includegraphics[width=0.4\textwidth]{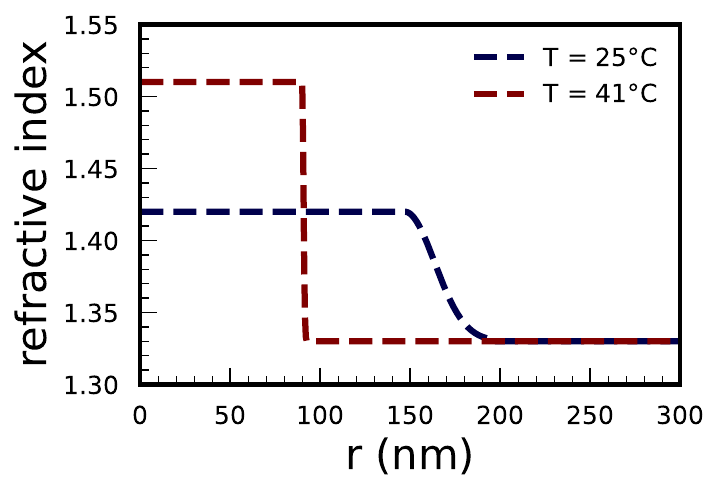}
\caption{Radial profile of the mnicrogel refractive index at the temperatures $T=25$°C and $T=41$°C, used in the electromagnetic full-wave simulations. \label{fig:figS9} \vspace{1.5cm}}
\end{figure}

Simulated extinction spectra of gold NPs with diameter $D=18.6$ nm randomly distributed on a spherical shell were obtained by numerically solving the full set of Maxwell’s equations using the finite element method (FEM) implemented in the commercial software COMSOL Multiphysics \cite{COMSOL}, utilizing the frequency domain solver (radio-frequency module).
The system's geometry for the simulations was created based on the configurations provided by the toy model described above. Specifically, a number $N_p$ of particles were positioned at the coordinates given by the toy model, corresponding to the temperatures $T=25$°C and $T=41$°C. The geometric domain was enclosed by a perfectly matched layer (PML) \cite{Yushanov2013} with external and internal radii equal to $R_\text{PML}^{(ext)}=2.5\,R$ and $R_\text{PML}^{(int)}=2.25\,R$ respectively, where $R$ is the external radius of the shell, according to the toy model (see Section S4). The extinction cross-section was obtained by adding the absorption and scattering cross-sections. The former was determined by integrating the power loss density over the NPs volume, and the latter was determined by integrating the Poynting vector over the spherical surface with the radius $R_\text{PML}^{(int)}=2.25\,R$. Both the contributions were normalized to the incident radiation. The incident field is a linearly-polarized plane wave $E_{\text{IN}} = \Re( E_0 e^{i (k x - \omega t)} )$, where $E_0$, $\omega$, and $k$ are the amplitude, frequency and wave-vector, respectively, and we defined as $x$ and $z$ the directions of the $k$-vector and of the polarization, respectively.
The microgel was modeled as a non-uniform background with a radial dielectric permittivity, assuming a constant value in the region $r<R$ and a Gaussian decay, as sketched in Fig.~\ref{fig:figS9}. The numerical values of the refractive index at the two temperature analysed were calculated by considering the permittivity of the pNIPAM microgels \cite{Brasse2018}, appropriately scaled based on the water volume fraction present in the polymer network in the swollen and shrunken states, as determined in previous studies \cite{Camerin2018,Bischofberger2015}. For the optical functions of gold NPs and water, we used the data reported in Rakić and Laurens, respectively \cite{Rakic1998,Laurens1999}. All domains in the simulation box were meshed using tetrahedral elements, maintaining a maximum element size below $\lambda/20$ for the outer domain, where $\lambda$ is the smallest wavelength used for the calculation of extinction spectra, and below $D/20$ for the NPs.
The maps of electric field enhancement, obtained at the wavelength $\lambda=630$ nm, where the coupling contribution is significant, are shown in Fig.~3B, allowing us for clearly visualize the resulting local field enhancement before and after the VPT. At $T=25$°C, the NPs on the microgel behave as isolated particles, whereas at $T=41$°C, the electric field is delocalized over adjacent NPs, as evidenced by the zoom of panel (iv), highlighting the activation of plasmon coupling.
The corresponding extinction spectra obtained from the full-wave simulations are reported in Fig.~3C and accurately reproduce the spectral features of the experimental data at both the analysed temperatures.

\bibliographystyle{apsrev4-2}
\bibliography{references}

\begin{thebibliography}{59}%
\makeatletter
\providecommand \@ifxundefined [1]{%
 \@ifx{#1\undefined}
}%
\providecommand \@ifnum [1]{%
 \ifnum #1\expandafter \@firstoftwo
 \else \expandafter \@secondoftwo
 \fi
}%
\providecommand \@ifx [1]{%
 \ifx #1\expandafter \@firstoftwo
 \else \expandafter \@secondoftwo
 \fi
}%
\providecommand \natexlab [1]{#1}%
\providecommand \enquote  [1]{``#1''}%
\providecommand \bibnamefont  [1]{#1}%
\providecommand \bibfnamefont [1]{#1}%
\providecommand \citenamefont [1]{#1}%
\providecommand \href@noop [0]{\@secondoftwo}%
\providecommand \href [0]{\begingroup \@sanitize@url \@href}%
\providecommand \@href[1]{\@@startlink{#1}\@@href}%
\providecommand \@@href[1]{\endgroup#1\@@endlink}%
\providecommand \@sanitize@url [0]{\catcode `\\12\catcode `\$12\catcode `\&12\catcode `\#12\catcode `\^12\catcode `\_12\catcode `\%12\relax}%
\providecommand \@@startlink[1]{}%
\providecommand \@@endlink[0]{}%
\providecommand \url  [0]{\begingroup\@sanitize@url \@url }%
\providecommand \@url [1]{\endgroup\@href {#1}{\urlprefix }}%
\providecommand \urlprefix  [0]{URL }%
\providecommand \Eprint [0]{\href }%
\providecommand \doibase [0]{https://doi.org/}%
\providecommand \selectlanguage [0]{\@gobble}%
\providecommand \bibinfo  [0]{\@secondoftwo}%
\providecommand \bibfield  [0]{\@secondoftwo}%
\providecommand \translation [1]{[#1]}%
\providecommand \BibitemOpen [0]{}%
\providecommand \bibitemStop [0]{}%
\providecommand \bibitemNoStop [0]{.\EOS\space}%
\providecommand \EOS [0]{\spacefactor3000\relax}%
\providecommand \BibitemShut  [1]{\csname bibitem#1\endcsname}%
\let\auto@bib@innerbib\@empty
\bibitem [{\citenamefont {Thomson}(1904)}]{thomson1904}%
  \BibitemOpen
  \bibfield  {author} {\bibinfo {author} {\bibfnamefont {J.~J.}\ \bibnamefont {Thomson}},\ }\href {https://doi.org/10.1080/14786440409463107} {\bibfield  {journal} {\bibinfo  {journal} {The London, Edinburgh, and Dublin Philosophical Magazine and Journal of Science}\ }\textbf {\bibinfo {volume} {7}},\ \bibinfo {pages} {237} (\bibinfo {year} {1904})}\BibitemShut {NoStop}%
\bibitem [{\citenamefont {Bowick}\ and\ \citenamefont {Giomi}(2009)}]{bowick2009}%
  \BibitemOpen
  \bibfield  {author} {\bibinfo {author} {\bibfnamefont {M.~J.}\ \bibnamefont {Bowick}}\ and\ \bibinfo {author} {\bibfnamefont {L.}~\bibnamefont {Giomi}},\ }\href {https://doi.org/10.1080/00018730903043166} {\bibfield  {journal} {\bibinfo  {journal} {Advances in Physics}\ }\textbf {\bibinfo {volume} {58}},\ \bibinfo {pages} {449} (\bibinfo {year} {2009})}\BibitemShut {NoStop}%
\bibitem [{\citenamefont {Vitelli}\ \emph {et~al.}(2006)\citenamefont {Vitelli}, \citenamefont {Lucks},\ and\ \citenamefont {Nelson}}]{vitelli2006}%
  \BibitemOpen
  \bibfield  {author} {\bibinfo {author} {\bibfnamefont {V.}~\bibnamefont {Vitelli}}, \bibinfo {author} {\bibfnamefont {J.~B.}\ \bibnamefont {Lucks}},\ and\ \bibinfo {author} {\bibfnamefont {D.~R.}\ \bibnamefont {Nelson}},\ }\href {https://doi.org/10.1073/pnas.060275510} {\bibfield  {journal} {\bibinfo  {journal} {Proceedings of the National Academy of Sciences}\ }\textbf {\bibinfo {volume} {103}},\ \bibinfo {pages} {12323} (\bibinfo {year} {2006})}\BibitemShut {NoStop}%
\bibitem [{\citenamefont {Mart{\'\i}n-Bravo}\ \emph {et~al.}(2021)\citenamefont {Mart{\'\i}n-Bravo}, \citenamefont {Llorente}, \citenamefont {Hern{\'a}ndez-Rojas},\ and\ \citenamefont {Wales}}]{martin2021}%
  \BibitemOpen
  \bibfield  {author} {\bibinfo {author} {\bibfnamefont {M.}~\bibnamefont {Mart{\'\i}n-Bravo}}, \bibinfo {author} {\bibfnamefont {J.~M.~G.}\ \bibnamefont {Llorente}}, \bibinfo {author} {\bibfnamefont {J.}~\bibnamefont {Hern{\'a}ndez-Rojas}},\ and\ \bibinfo {author} {\bibfnamefont {D.~J.}\ \bibnamefont {Wales}},\ }\href {https://doi.org/10.1021/acsnano.1c04952} {\bibfield  {journal} {\bibinfo  {journal} {ACS nano}\ }\textbf {\bibinfo {volume} {15}},\ \bibinfo {pages} {14873} (\bibinfo {year} {2021})}\BibitemShut {NoStop}%
\bibitem [{\citenamefont {Fantoni}\ \emph {et~al.}(2012)\citenamefont {Fantoni}, \citenamefont {Salari},\ and\ \citenamefont {Klumperman}}]{fantoni2012}%
  \BibitemOpen
  \bibfield  {author} {\bibinfo {author} {\bibfnamefont {R.}~\bibnamefont {Fantoni}}, \bibinfo {author} {\bibfnamefont {J.~W.}\ \bibnamefont {Salari}},\ and\ \bibinfo {author} {\bibfnamefont {B.}~\bibnamefont {Klumperman}},\ }\href {https://doi.org/10.1103/physreve.85.061404} {\bibfield  {journal} {\bibinfo  {journal} {Physical Review E—Statistical, Nonlinear, and Soft Matter Physics}\ }\textbf {\bibinfo {volume} {85}},\ \bibinfo {pages} {061404} (\bibinfo {year} {2012})}\BibitemShut {NoStop}%
\bibitem [{\citenamefont {Bo{\v{z}}i{\v{c}}}\ and\ \citenamefont {{\v{C}}opar}(2019)}]{bozic2019}%
  \BibitemOpen
  \bibfield  {author} {\bibinfo {author} {\bibfnamefont {A.~L.}\ \bibnamefont {Bo{\v{z}}i{\v{c}}}}\ and\ \bibinfo {author} {\bibfnamefont {S.}~\bibnamefont {{\v{C}}opar}},\ }\href {https://doi.org/10.1103/physreve.99.032601} {\bibfield  {journal} {\bibinfo  {journal} {Physical Review E}\ }\textbf {\bibinfo {volume} {99}},\ \bibinfo {pages} {032601} (\bibinfo {year} {2019})}\BibitemShut {NoStop}%
\bibitem [{\citenamefont {Javidpour}\ \emph {et~al.}(2021)\citenamefont {Javidpour}, \citenamefont {Bo{\v{z}}i{\v{c}}}, \citenamefont {Naji},\ and\ \citenamefont {Podgornik}}]{javidpour2021}%
  \BibitemOpen
  \bibfield  {author} {\bibinfo {author} {\bibfnamefont {L.}~\bibnamefont {Javidpour}}, \bibinfo {author} {\bibfnamefont {A.}~\bibnamefont {Bo{\v{z}}i{\v{c}}}}, \bibinfo {author} {\bibfnamefont {A.}~\bibnamefont {Naji}},\ and\ \bibinfo {author} {\bibfnamefont {R.}~\bibnamefont {Podgornik}},\ }\href {https://doi.org/10.1039/d1sm00232e} {\bibfield  {journal} {\bibinfo  {journal} {Soft Matter}\ }\textbf {\bibinfo {volume} {17}},\ \bibinfo {pages} {4296} (\bibinfo {year} {2021})}\BibitemShut {NoStop}%
\bibitem [{\citenamefont {Meyra}\ \emph {et~al.}(2019)\citenamefont {Meyra}, \citenamefont {Zarragoicoechea}, \citenamefont {Maltz}, \citenamefont {Lomba},\ and\ \citenamefont {Torquato}}]{meyra2019}%
  \BibitemOpen
  \bibfield  {author} {\bibinfo {author} {\bibfnamefont {A.~G.}\ \bibnamefont {Meyra}}, \bibinfo {author} {\bibfnamefont {G.~J.}\ \bibnamefont {Zarragoicoechea}}, \bibinfo {author} {\bibfnamefont {A.~L.}\ \bibnamefont {Maltz}}, \bibinfo {author} {\bibfnamefont {E.}~\bibnamefont {Lomba}},\ and\ \bibinfo {author} {\bibfnamefont {S.}~\bibnamefont {Torquato}},\ }\href {https://doi.org/10.1103/physreve.100.022107} {\bibfield  {journal} {\bibinfo  {journal} {Physical Review E}\ }\textbf {\bibinfo {volume} {100}},\ \bibinfo {pages} {022107} (\bibinfo {year} {2019})}\BibitemShut {NoStop}%
\bibitem [{\citenamefont {Carenza}\ \emph {et~al.}(2022)\citenamefont {Carenza}, \citenamefont {Gonnella}, \citenamefont {Marenduzzo}, \citenamefont {Negro},\ and\ \citenamefont {Orlandini}}]{carenza2022}%
  \BibitemOpen
  \bibfield  {author} {\bibinfo {author} {\bibfnamefont {L.~N.}\ \bibnamefont {Carenza}}, \bibinfo {author} {\bibfnamefont {G.}~\bibnamefont {Gonnella}}, \bibinfo {author} {\bibfnamefont {D.}~\bibnamefont {Marenduzzo}}, \bibinfo {author} {\bibfnamefont {G.}~\bibnamefont {Negro}},\ and\ \bibinfo {author} {\bibfnamefont {E.}~\bibnamefont {Orlandini}},\ }\href {https://doi.org/10.1103/physrevlett.128.027801} {\bibfield  {journal} {\bibinfo  {journal} {Physical Review Letters}\ }\textbf {\bibinfo {volume} {128}},\ \bibinfo {pages} {027801} (\bibinfo {year} {2022})}\BibitemShut {NoStop}%
\bibitem [{\citenamefont {Viveros-M{\'e}ndez}\ \emph {et~al.}(2008)\citenamefont {Viveros-M{\'e}ndez}, \citenamefont {M{\'e}ndez-Alcaraz},\ and\ \citenamefont {Gonz{\'a}lez-Mozuelos}}]{viveros2008}%
  \BibitemOpen
  \bibfield  {author} {\bibinfo {author} {\bibfnamefont {P.}~\bibnamefont {Viveros-M{\'e}ndez}}, \bibinfo {author} {\bibfnamefont {J.}~\bibnamefont {M{\'e}ndez-Alcaraz}},\ and\ \bibinfo {author} {\bibfnamefont {P.}~\bibnamefont {Gonz{\'a}lez-Mozuelos}},\ }\bibfield  {journal} {\bibinfo  {journal} {The Journal of Chemical Physics}\ }\textbf {\bibinfo {volume} {128}},\ \href {https://doi.org/10.1063/1.2816558} {10.1063/1.2816558} (\bibinfo {year} {2008})\BibitemShut {NoStop}%
\bibitem [{\citenamefont {Bausch}\ \emph {et~al.}(2003)\citenamefont {Bausch}, \citenamefont {Bowick}, \citenamefont {Cacciuto}, \citenamefont {Dinsmore}, \citenamefont {Hsu}, \citenamefont {Nelson}, \citenamefont {Nikolaides}, \citenamefont {Travesset},\ and\ \citenamefont {Weitz}}]{bausch2003}%
  \BibitemOpen
  \bibfield  {author} {\bibinfo {author} {\bibfnamefont {A.}~\bibnamefont {Bausch}}, \bibinfo {author} {\bibfnamefont {M.~J.}\ \bibnamefont {Bowick}}, \bibinfo {author} {\bibfnamefont {A.}~\bibnamefont {Cacciuto}}, \bibinfo {author} {\bibfnamefont {A.}~\bibnamefont {Dinsmore}}, \bibinfo {author} {\bibfnamefont {M.}~\bibnamefont {Hsu}}, \bibinfo {author} {\bibfnamefont {D.}~\bibnamefont {Nelson}}, \bibinfo {author} {\bibfnamefont {M.}~\bibnamefont {Nikolaides}}, \bibinfo {author} {\bibfnamefont {A.}~\bibnamefont {Travesset}},\ and\ \bibinfo {author} {\bibfnamefont {D.}~\bibnamefont {Weitz}},\ }\href {https://doi.org/10.1126/science.1081160} {\bibfield  {journal} {\bibinfo  {journal} {Science}\ }\textbf {\bibinfo {volume} {299}},\ \bibinfo {pages} {1716} (\bibinfo {year} {2003})}\BibitemShut {NoStop}%
\bibitem [{\citenamefont {Sun}\ \emph {et~al.}(2025)\citenamefont {Sun}, \citenamefont {Zhang}, \citenamefont {Plummer}, \citenamefont {Martin}, \citenamefont {Tanjeem}, \citenamefont {Nelson},\ and\ \citenamefont {Manoharan}}]{sun2025}%
  \BibitemOpen
  \bibfield  {author} {\bibinfo {author} {\bibfnamefont {J.~H.}\ \bibnamefont {Sun}}, \bibinfo {author} {\bibfnamefont {G.~H.}\ \bibnamefont {Zhang}}, \bibinfo {author} {\bibfnamefont {A.}~\bibnamefont {Plummer}}, \bibinfo {author} {\bibfnamefont {C.}~\bibnamefont {Martin}}, \bibinfo {author} {\bibfnamefont {N.}~\bibnamefont {Tanjeem}}, \bibinfo {author} {\bibfnamefont {D.~R.}\ \bibnamefont {Nelson}},\ and\ \bibinfo {author} {\bibfnamefont {V.~N.}\ \bibnamefont {Manoharan}},\ }\href {https://doi.org/10.1103/physrevlett.134.018201} {\bibfield  {journal} {\bibinfo  {journal} {Physical Review Letters}\ }\textbf {\bibinfo {volume} {134}},\ \bibinfo {pages} {018201} (\bibinfo {year} {2025})}\BibitemShut {NoStop}%
\bibitem [{\citenamefont {Singh}\ \emph {et~al.}(2022)\citenamefont {Singh}, \citenamefont {Sood},\ and\ \citenamefont {Ganapathy}}]{singh2022}%
  \BibitemOpen
  \bibfield  {author} {\bibinfo {author} {\bibfnamefont {N.}~\bibnamefont {Singh}}, \bibinfo {author} {\bibfnamefont {A.}~\bibnamefont {Sood}},\ and\ \bibinfo {author} {\bibfnamefont {R.}~\bibnamefont {Ganapathy}},\ }\href {https://doi.org/10.1073/pnas.2206470119} {\bibfield  {journal} {\bibinfo  {journal} {Proceedings of the National Academy of Sciences}\ }\textbf {\bibinfo {volume} {119}},\ \bibinfo {pages} {e2206470119} (\bibinfo {year} {2022})}\BibitemShut {NoStop}%
\bibitem [{\citenamefont {Guerra}\ \emph {et~al.}(2018)\citenamefont {Guerra}, \citenamefont {Kelleher}, \citenamefont {Hollingsworth},\ and\ \citenamefont {Chaikin}}]{guerra2018}%
  \BibitemOpen
  \bibfield  {author} {\bibinfo {author} {\bibfnamefont {R.~E.}\ \bibnamefont {Guerra}}, \bibinfo {author} {\bibfnamefont {C.~P.}\ \bibnamefont {Kelleher}}, \bibinfo {author} {\bibfnamefont {A.~D.}\ \bibnamefont {Hollingsworth}},\ and\ \bibinfo {author} {\bibfnamefont {P.~M.}\ \bibnamefont {Chaikin}},\ }\href {https://doi.org/10.1038/nature25468} {\bibfield  {journal} {\bibinfo  {journal} {Nature}\ }\textbf {\bibinfo {volume} {554}},\ \bibinfo {pages} {346} (\bibinfo {year} {2018})}\BibitemShut {NoStop}%
\bibitem [{\citenamefont {Silvera~Batista}\ \emph {et~al.}(2015)\citenamefont {Silvera~Batista}, \citenamefont {Larson},\ and\ \citenamefont {Kotov}}]{silvera2015}%
  \BibitemOpen
  \bibfield  {author} {\bibinfo {author} {\bibfnamefont {C.~A.}\ \bibnamefont {Silvera~Batista}}, \bibinfo {author} {\bibfnamefont {R.~G.}\ \bibnamefont {Larson}},\ and\ \bibinfo {author} {\bibfnamefont {N.~A.}\ \bibnamefont {Kotov}},\ }\href {https://doi.org/10.1126/science.1242477} {\bibfield  {journal} {\bibinfo  {journal} {Science}\ }\textbf {\bibinfo {volume} {350}},\ \bibinfo {pages} {1242477} (\bibinfo {year} {2015})}\BibitemShut {NoStop}%
\bibitem [{\citenamefont {Fernandez-Nieves}\ \emph {et~al.}(2011)\citenamefont {Fernandez-Nieves}, \citenamefont {Wyss}, \citenamefont {Mattsson},\ and\ \citenamefont {Weitz}}]{fernandez2011}%
  \BibitemOpen
  \bibfield  {author} {\bibinfo {author} {\bibfnamefont {A.}~\bibnamefont {Fernandez-Nieves}}, \bibinfo {author} {\bibfnamefont {H.}~\bibnamefont {Wyss}}, \bibinfo {author} {\bibfnamefont {J.}~\bibnamefont {Mattsson}},\ and\ \bibinfo {author} {\bibfnamefont {D.~A.}\ \bibnamefont {Weitz}},\ }\href@noop {} {\emph {\bibinfo {title} {Microgel Suspensions: Fundamentals and Applications}}}\ (\bibinfo  {publisher} {John Wiley \& Sons},\ \bibinfo {year} {2011})\BibitemShut {NoStop}%
\bibitem [{\citenamefont {Kelleher}\ \emph {et~al.}(2017)\citenamefont {Kelleher}, \citenamefont {Guerra}, \citenamefont {Hollingsworth},\ and\ \citenamefont {Chaikin}}]{kelleher2017}%
  \BibitemOpen
  \bibfield  {author} {\bibinfo {author} {\bibfnamefont {C.~P.}\ \bibnamefont {Kelleher}}, \bibinfo {author} {\bibfnamefont {R.~E.}\ \bibnamefont {Guerra}}, \bibinfo {author} {\bibfnamefont {A.~D.}\ \bibnamefont {Hollingsworth}},\ and\ \bibinfo {author} {\bibfnamefont {P.~M.}\ \bibnamefont {Chaikin}},\ }\href {https://doi.org/10.1103/physreve.95.022602} {\bibfield  {journal} {\bibinfo  {journal} {Physical Review E}\ }\textbf {\bibinfo {volume} {95}},\ \bibinfo {pages} {022602} (\bibinfo {year} {2017})}\BibitemShut {NoStop}%
\bibitem [{\citenamefont {Gawlitza}\ \emph {et~al.}(2013)\citenamefont {Gawlitza}, \citenamefont {Turner}, \citenamefont {Polzer}, \citenamefont {Wellert}, \citenamefont {Karg}, \citenamefont {Mulvaney},\ and\ \citenamefont {von Klitzing}}]{gawlitza2013}%
  \BibitemOpen
  \bibfield  {author} {\bibinfo {author} {\bibfnamefont {K.}~\bibnamefont {Gawlitza}}, \bibinfo {author} {\bibfnamefont {S.~T.}\ \bibnamefont {Turner}}, \bibinfo {author} {\bibfnamefont {F.}~\bibnamefont {Polzer}}, \bibinfo {author} {\bibfnamefont {S.}~\bibnamefont {Wellert}}, \bibinfo {author} {\bibfnamefont {M.}~\bibnamefont {Karg}}, \bibinfo {author} {\bibfnamefont {P.}~\bibnamefont {Mulvaney}},\ and\ \bibinfo {author} {\bibfnamefont {R.}~\bibnamefont {von Klitzing}},\ }\href {https://doi.org/10.1039/c3cp51578h} {\bibfield  {journal} {\bibinfo  {journal} {Physical Chemistry Chemical Physics}\ }\textbf {\bibinfo {volume} {15}},\ \bibinfo {pages} {15623} (\bibinfo {year} {2013})}\BibitemShut {NoStop}%
\bibitem [{\citenamefont {Suzuki}\ \emph {et~al.}(2014)\citenamefont {Suzuki}, \citenamefont {Nagase}, \citenamefont {Kureha},\ and\ \citenamefont {Sato}}]{suzuki2014}%
  \BibitemOpen
  \bibfield  {author} {\bibinfo {author} {\bibfnamefont {D.}~\bibnamefont {Suzuki}}, \bibinfo {author} {\bibfnamefont {Y.}~\bibnamefont {Nagase}}, \bibinfo {author} {\bibfnamefont {T.}~\bibnamefont {Kureha}},\ and\ \bibinfo {author} {\bibfnamefont {T.}~\bibnamefont {Sato}},\ }\href {https://doi.org/10.1021/jp410983x} {\bibfield  {journal} {\bibinfo  {journal} {The Journal of Physical Chemistry B}\ }\textbf {\bibinfo {volume} {118}},\ \bibinfo {pages} {2194} (\bibinfo {year} {2014})}\BibitemShut {NoStop}%
\bibitem [{\citenamefont {Choe}\ \emph {et~al.}(2018)\citenamefont {Choe}, \citenamefont {Yeom}, \citenamefont {Shanker}, \citenamefont {Kim}, \citenamefont {Kang},\ and\ \citenamefont {Ko}}]{choe2018}%
  \BibitemOpen
  \bibfield  {author} {\bibinfo {author} {\bibfnamefont {A.}~\bibnamefont {Choe}}, \bibinfo {author} {\bibfnamefont {J.}~\bibnamefont {Yeom}}, \bibinfo {author} {\bibfnamefont {R.}~\bibnamefont {Shanker}}, \bibinfo {author} {\bibfnamefont {M.~P.}\ \bibnamefont {Kim}}, \bibinfo {author} {\bibfnamefont {S.}~\bibnamefont {Kang}},\ and\ \bibinfo {author} {\bibfnamefont {H.}~\bibnamefont {Ko}},\ }\href {https://doi.org/10.1038/s41427-018-0086-6} {\bibfield  {journal} {\bibinfo  {journal} {NPG Asia Materials}\ }\textbf {\bibinfo {volume} {10}},\ \bibinfo {pages} {912} (\bibinfo {year} {2018})}\BibitemShut {NoStop}%
\bibitem [{\citenamefont {Sabadasch}\ \emph {et~al.}(2020)\citenamefont {Sabadasch}, \citenamefont {Wiehemeier}, \citenamefont {Kottke},\ and\ \citenamefont {Hellweg}}]{sabadasch2020}%
  \BibitemOpen
  \bibfield  {author} {\bibinfo {author} {\bibfnamefont {V.}~\bibnamefont {Sabadasch}}, \bibinfo {author} {\bibfnamefont {L.}~\bibnamefont {Wiehemeier}}, \bibinfo {author} {\bibfnamefont {T.}~\bibnamefont {Kottke}},\ and\ \bibinfo {author} {\bibfnamefont {T.}~\bibnamefont {Hellweg}},\ }\href {https://doi.org/10.1039/d0sm00433b} {\bibfield  {journal} {\bibinfo  {journal} {Soft Matter}\ }\textbf {\bibinfo {volume} {16}},\ \bibinfo {pages} {5422} (\bibinfo {year} {2020})}\BibitemShut {NoStop}%
\bibitem [{\citenamefont {Arif}\ \emph {et~al.}(2021)\citenamefont {Arif}, \citenamefont {Farooqi}, \citenamefont {Irfan},\ and\ \citenamefont {Begum}}]{arif2021}%
  \BibitemOpen
  \bibfield  {author} {\bibinfo {author} {\bibfnamefont {M.}~\bibnamefont {Arif}}, \bibinfo {author} {\bibfnamefont {Z.~H.}\ \bibnamefont {Farooqi}}, \bibinfo {author} {\bibfnamefont {A.}~\bibnamefont {Irfan}},\ and\ \bibinfo {author} {\bibfnamefont {R.}~\bibnamefont {Begum}},\ }\href {https://doi.org/10.1016/j.molliq.2021.116270} {\bibfield  {journal} {\bibinfo  {journal} {Journal of Molecular Liquids}\ }\textbf {\bibinfo {volume} {336}},\ \bibinfo {pages} {116270} (\bibinfo {year} {2021})}\BibitemShut {NoStop}%
\bibitem [{\citenamefont {Diehl}\ \emph {et~al.}(2022)\citenamefont {Diehl}, \citenamefont {Hageneder}, \citenamefont {Fossati}, \citenamefont {Auer}, \citenamefont {Dostalek},\ and\ \citenamefont {Jonas}}]{diehl2022}%
  \BibitemOpen
  \bibfield  {author} {\bibinfo {author} {\bibfnamefont {F.}~\bibnamefont {Diehl}}, \bibinfo {author} {\bibfnamefont {S.}~\bibnamefont {Hageneder}}, \bibinfo {author} {\bibfnamefont {S.}~\bibnamefont {Fossati}}, \bibinfo {author} {\bibfnamefont {S.~K.}\ \bibnamefont {Auer}}, \bibinfo {author} {\bibfnamefont {J.}~\bibnamefont {Dostalek}},\ and\ \bibinfo {author} {\bibfnamefont {U.}~\bibnamefont {Jonas}},\ }\href {https://doi.org/10.1039/d1cs01083b} {\bibfield  {journal} {\bibinfo  {journal} {Chemical Society Reviews}\ }\textbf {\bibinfo {volume} {51}},\ \bibinfo {pages} {3926} (\bibinfo {year} {2022})}\BibitemShut {NoStop}%
\bibitem [{\citenamefont {Halas}\ \emph {et~al.}(2011)\citenamefont {Halas}, \citenamefont {Lal}, \citenamefont {Chang}, \citenamefont {Link},\ and\ \citenamefont {Nordlander}}]{halas2011}%
  \BibitemOpen
  \bibfield  {author} {\bibinfo {author} {\bibfnamefont {N.~J.}\ \bibnamefont {Halas}}, \bibinfo {author} {\bibfnamefont {S.}~\bibnamefont {Lal}}, \bibinfo {author} {\bibfnamefont {W.-S.}\ \bibnamefont {Chang}}, \bibinfo {author} {\bibfnamefont {S.}~\bibnamefont {Link}},\ and\ \bibinfo {author} {\bibfnamefont {P.}~\bibnamefont {Nordlander}},\ }\href {https://doi.org/10.1021/cr200061k} {\bibfield  {journal} {\bibinfo  {journal} {Chemical reviews}\ }\textbf {\bibinfo {volume} {111}},\ \bibinfo {pages} {3913} (\bibinfo {year} {2011})}\BibitemShut {NoStop}%
\bibitem [{\citenamefont {Li}\ \emph {et~al.}(2020)\citenamefont {Li}, \citenamefont {Wang},\ and\ \citenamefont {Yin}}]{li2020}%
  \BibitemOpen
  \bibfield  {author} {\bibinfo {author} {\bibfnamefont {Z.}~\bibnamefont {Li}}, \bibinfo {author} {\bibfnamefont {W.}~\bibnamefont {Wang}},\ and\ \bibinfo {author} {\bibfnamefont {Y.}~\bibnamefont {Yin}},\ }\href {https://doi.org/10.1016/j.trechm.2020.03.008} {\bibfield  {journal} {\bibinfo  {journal} {Trends in Chemistry}\ }\textbf {\bibinfo {volume} {2}},\ \bibinfo {pages} {593} (\bibinfo {year} {2020})}\BibitemShut {NoStop}%
\bibitem [{\citenamefont {Capocefalo}\ \emph {et~al.}(2022)\citenamefont {Capocefalo}, \citenamefont {Bizien}, \citenamefont {Sennato}, \citenamefont {Ghofraniha}, \citenamefont {Bordi},\ and\ \citenamefont {Brasili}}]{capocefalo2022}%
  \BibitemOpen
  \bibfield  {author} {\bibinfo {author} {\bibfnamefont {A.}~\bibnamefont {Capocefalo}}, \bibinfo {author} {\bibfnamefont {T.}~\bibnamefont {Bizien}}, \bibinfo {author} {\bibfnamefont {S.}~\bibnamefont {Sennato}}, \bibinfo {author} {\bibfnamefont {N.}~\bibnamefont {Ghofraniha}}, \bibinfo {author} {\bibfnamefont {F.}~\bibnamefont {Bordi}},\ and\ \bibinfo {author} {\bibfnamefont {F.}~\bibnamefont {Brasili}},\ }\href {https://doi.org/10.3390/nano12091529} {\bibfield  {journal} {\bibinfo  {journal} {Nanomaterials}\ }\textbf {\bibinfo {volume} {12}},\ \bibinfo {pages} {1529} (\bibinfo {year} {2022})}\BibitemShut {NoStop}%
\bibitem [{\citenamefont {Liu}\ \emph {et~al.}(2014)\citenamefont {Liu}, \citenamefont {Yang}, \citenamefont {Meng}, \citenamefont {Sun}, \citenamefont {Wang}, \citenamefont {Yang}, \citenamefont {Liu},\ and\ \citenamefont {Tian}}]{liu2014}%
  \BibitemOpen
  \bibfield  {author} {\bibinfo {author} {\bibfnamefont {H.}~\bibnamefont {Liu}}, \bibinfo {author} {\bibfnamefont {Z.}~\bibnamefont {Yang}}, \bibinfo {author} {\bibfnamefont {L.}~\bibnamefont {Meng}}, \bibinfo {author} {\bibfnamefont {Y.}~\bibnamefont {Sun}}, \bibinfo {author} {\bibfnamefont {J.}~\bibnamefont {Wang}}, \bibinfo {author} {\bibfnamefont {L.}~\bibnamefont {Yang}}, \bibinfo {author} {\bibfnamefont {J.}~\bibnamefont {Liu}},\ and\ \bibinfo {author} {\bibfnamefont {Z.}~\bibnamefont {Tian}},\ }\href {https://doi.org/10.1021/ja501951v} {\bibfield  {journal} {\bibinfo  {journal} {Journal of the American Chemical Society}\ }\textbf {\bibinfo {volume} {136}},\ \bibinfo {pages} {5332} (\bibinfo {year} {2014})}\BibitemShut {NoStop}%
\bibitem [{\citenamefont {Caprara}\ \emph {et~al.}(2020)\citenamefont {Caprara}, \citenamefont {Ripanti}, \citenamefont {Capocefalo}, \citenamefont {Sarra}, \citenamefont {Brasili}, \citenamefont {Petrillo}, \citenamefont {Fasolato},\ and\ \citenamefont {Postorino}}]{caprara2020}%
  \BibitemOpen
  \bibfield  {author} {\bibinfo {author} {\bibfnamefont {D.}~\bibnamefont {Caprara}}, \bibinfo {author} {\bibfnamefont {F.}~\bibnamefont {Ripanti}}, \bibinfo {author} {\bibfnamefont {A.}~\bibnamefont {Capocefalo}}, \bibinfo {author} {\bibfnamefont {A.}~\bibnamefont {Sarra}}, \bibinfo {author} {\bibfnamefont {F.}~\bibnamefont {Brasili}}, \bibinfo {author} {\bibfnamefont {C.}~\bibnamefont {Petrillo}}, \bibinfo {author} {\bibfnamefont {C.}~\bibnamefont {Fasolato}},\ and\ \bibinfo {author} {\bibfnamefont {P.}~\bibnamefont {Postorino}},\ }\href {https://doi.org/10.1016/j.colsurfa.2019.124399} {\bibfield  {journal} {\bibinfo  {journal} {Colloids and Surfaces A: Physicochemical and Engineering Aspects}\ }\textbf {\bibinfo {volume} {589}},\ \bibinfo {pages} {124399} (\bibinfo {year} {2020})}\BibitemShut {NoStop}%
\bibitem [{\citenamefont {Gnan}\ \emph {et~al.}(2017)\citenamefont {Gnan}, \citenamefont {Rovigatti}, \citenamefont {Bergman},\ and\ \citenamefont {Zaccarelli}}]{gnan2017}%
  \BibitemOpen
  \bibfield  {author} {\bibinfo {author} {\bibfnamefont {N.}~\bibnamefont {Gnan}}, \bibinfo {author} {\bibfnamefont {L.}~\bibnamefont {Rovigatti}}, \bibinfo {author} {\bibfnamefont {M.}~\bibnamefont {Bergman}},\ and\ \bibinfo {author} {\bibfnamefont {E.}~\bibnamefont {Zaccarelli}},\ }\href {https://doi.org/10.1021/acs.macromol.7b01600} {\bibfield  {journal} {\bibinfo  {journal} {Macromolecules}\ }\textbf {\bibinfo {volume} {50}},\ \bibinfo {pages} {8777} (\bibinfo {year} {2017})}\BibitemShut {NoStop}%
\bibitem [{\citenamefont {Ninarello}\ \emph {et~al.}(2019)\citenamefont {Ninarello}, \citenamefont {Crassous}, \citenamefont {Paloli}, \citenamefont {Camerin}, \citenamefont {Gnan}, \citenamefont {Rovigatti}, \citenamefont {Schurtenberger},\ and\ \citenamefont {Zaccarelli}}]{ninarello2019}%
  \BibitemOpen
  \bibfield  {author} {\bibinfo {author} {\bibfnamefont {A.}~\bibnamefont {Ninarello}}, \bibinfo {author} {\bibfnamefont {J.~J.}\ \bibnamefont {Crassous}}, \bibinfo {author} {\bibfnamefont {D.}~\bibnamefont {Paloli}}, \bibinfo {author} {\bibfnamefont {F.}~\bibnamefont {Camerin}}, \bibinfo {author} {\bibfnamefont {N.}~\bibnamefont {Gnan}}, \bibinfo {author} {\bibfnamefont {L.}~\bibnamefont {Rovigatti}}, \bibinfo {author} {\bibfnamefont {P.}~\bibnamefont {Schurtenberger}},\ and\ \bibinfo {author} {\bibfnamefont {E.}~\bibnamefont {Zaccarelli}},\ }\href {https://doi.org/10.1021/acs.macromol.9b01122} {\bibfield  {journal} {\bibinfo  {journal} {Macromolecules}\ }\textbf {\bibinfo {volume} {52}},\ \bibinfo {pages} {7584} (\bibinfo {year} {2019})}\BibitemShut {NoStop}%
\bibitem [{\citenamefont {Del~Monte}\ \emph {et~al.}(2019)\citenamefont {Del~Monte}, \citenamefont {Ninarello}, \citenamefont {Camerin}, \citenamefont {Rovigatti}, \citenamefont {Gnan},\ and\ \citenamefont {Zaccarelli}}]{del2019}%
  \BibitemOpen
  \bibfield  {author} {\bibinfo {author} {\bibfnamefont {G.}~\bibnamefont {Del~Monte}}, \bibinfo {author} {\bibfnamefont {A.}~\bibnamefont {Ninarello}}, \bibinfo {author} {\bibfnamefont {F.}~\bibnamefont {Camerin}}, \bibinfo {author} {\bibfnamefont {L.}~\bibnamefont {Rovigatti}}, \bibinfo {author} {\bibfnamefont {N.}~\bibnamefont {Gnan}},\ and\ \bibinfo {author} {\bibfnamefont {E.}~\bibnamefont {Zaccarelli}},\ }\href {https://doi.org/10.1039/c9sm01253b} {\bibfield  {journal} {\bibinfo  {journal} {Soft Matter}\ }\textbf {\bibinfo {volume} {15}},\ \bibinfo {pages} {8113} (\bibinfo {year} {2019})}\BibitemShut {NoStop}%
\bibitem [{\citenamefont {Brasili}\ \emph {et~al.}(2023)\citenamefont {Brasili}, \citenamefont {Del~Monte}, \citenamefont {Capocefalo}, \citenamefont {Chauveau}, \citenamefont {Buratti}, \citenamefont {Casciardi}, \citenamefont {Truzzolillo}, \citenamefont {Sennato},\ and\ \citenamefont {Zaccarelli}}]{brasili2023}%
  \BibitemOpen
  \bibfield  {author} {\bibinfo {author} {\bibfnamefont {F.}~\bibnamefont {Brasili}}, \bibinfo {author} {\bibfnamefont {G.}~\bibnamefont {Del~Monte}}, \bibinfo {author} {\bibfnamefont {A.}~\bibnamefont {Capocefalo}}, \bibinfo {author} {\bibfnamefont {E.}~\bibnamefont {Chauveau}}, \bibinfo {author} {\bibfnamefont {E.}~\bibnamefont {Buratti}}, \bibinfo {author} {\bibfnamefont {S.}~\bibnamefont {Casciardi}}, \bibinfo {author} {\bibfnamefont {D.}~\bibnamefont {Truzzolillo}}, \bibinfo {author} {\bibfnamefont {S.}~\bibnamefont {Sennato}},\ and\ \bibinfo {author} {\bibfnamefont {E.}~\bibnamefont {Zaccarelli}},\ }\href {https://doi.org/10.1021/acsami.3c14608} {\bibfield  {journal} {\bibinfo  {journal} {ACS Applied Materials \& Interfaces}\ }\textbf {\bibinfo {volume} {15}},\ \bibinfo {pages} {58770} (\bibinfo {year} {2023})}\BibitemShut {NoStop}%
\bibitem [{\citenamefont {Soddemann}\ \emph {et~al.}(2001)\citenamefont {Soddemann}, \citenamefont {D{\"u}nweg},\ and\ \citenamefont {Kremer}}]{soddemann2001}%
  \BibitemOpen
  \bibfield  {author} {\bibinfo {author} {\bibfnamefont {T.}~\bibnamefont {Soddemann}}, \bibinfo {author} {\bibfnamefont {B.}~\bibnamefont {D{\"u}nweg}},\ and\ \bibinfo {author} {\bibfnamefont {K.}~\bibnamefont {Kremer}},\ }\href {https://doi.org/10.1007/s10189-001-8054-4} {\bibfield  {journal} {\bibinfo  {journal} {The European Physical Journal E}\ }\textbf {\bibinfo {volume} {6}},\ \bibinfo {pages} {409} (\bibinfo {year} {2001})}\BibitemShut {NoStop}%
\bibitem [{\citenamefont {Jain}\ \emph {et~al.}(2007)\citenamefont {Jain}, \citenamefont {Huang},\ and\ \citenamefont {El-Sayed}}]{jain2007}%
  \BibitemOpen
  \bibfield  {author} {\bibinfo {author} {\bibfnamefont {P.~K.}\ \bibnamefont {Jain}}, \bibinfo {author} {\bibfnamefont {W.}~\bibnamefont {Huang}},\ and\ \bibinfo {author} {\bibfnamefont {M.~A.}\ \bibnamefont {El-Sayed}},\ }\href {https://doi.org/10.1021/nl071008a} {\bibfield  {journal} {\bibinfo  {journal} {Nano letters}\ }\textbf {\bibinfo {volume} {7}},\ \bibinfo {pages} {2080} (\bibinfo {year} {2007})}\BibitemShut {NoStop}%
\bibitem [{\citenamefont {S{\"o}nnichsen}\ \emph {et~al.}(2005)\citenamefont {S{\"o}nnichsen}, \citenamefont {Reinhard}, \citenamefont {Liphardt},\ and\ \citenamefont {Alivisatos}}]{sonnichsen2005}%
  \BibitemOpen
  \bibfield  {author} {\bibinfo {author} {\bibfnamefont {C.}~\bibnamefont {S{\"o}nnichsen}}, \bibinfo {author} {\bibfnamefont {B.~M.}\ \bibnamefont {Reinhard}}, \bibinfo {author} {\bibfnamefont {J.}~\bibnamefont {Liphardt}},\ and\ \bibinfo {author} {\bibfnamefont {A.~P.}\ \bibnamefont {Alivisatos}},\ }\href {https://doi.org/10.1038/nbt1100} {\bibfield  {journal} {\bibinfo  {journal} {Nature biotechnology}\ }\textbf {\bibinfo {volume} {23}},\ \bibinfo {pages} {741} (\bibinfo {year} {2005})}\BibitemShut {NoStop}%
\bibitem [{\citenamefont {Liu}\ \emph {et~al.}(2015)\citenamefont {Liu}, \citenamefont {Wang}, \citenamefont {Zhou},\ and\ \citenamefont {Liu}}]{liu2015electrostatic}%
  \BibitemOpen
  \bibfield  {author} {\bibinfo {author} {\bibfnamefont {G.}~\bibnamefont {Liu}}, \bibinfo {author} {\bibfnamefont {D.}~\bibnamefont {Wang}}, \bibinfo {author} {\bibfnamefont {F.}~\bibnamefont {Zhou}},\ and\ \bibinfo {author} {\bibfnamefont {W.}~\bibnamefont {Liu}},\ }\href@noop {} {\bibfield  {journal} {\bibinfo  {journal} {Small}\ }\textbf {\bibinfo {volume} {11}},\ \bibinfo {pages} {2807} (\bibinfo {year} {2015})}\BibitemShut {NoStop}%
\bibitem [{\citenamefont {Chang}\ \emph {et~al.}(2023)\citenamefont {Chang}, \citenamefont {Yan}, \citenamefont {Zhang}, \citenamefont {Xia}, \citenamefont {Chen}, \citenamefont {Lei},\ and\ \citenamefont {Shi}}]{chang2023synergistic}%
  \BibitemOpen
  \bibfield  {author} {\bibinfo {author} {\bibfnamefont {K.}~\bibnamefont {Chang}}, \bibinfo {author} {\bibfnamefont {Y.}~\bibnamefont {Yan}}, \bibinfo {author} {\bibfnamefont {D.}~\bibnamefont {Zhang}}, \bibinfo {author} {\bibfnamefont {Y.}~\bibnamefont {Xia}}, \bibinfo {author} {\bibfnamefont {X.}~\bibnamefont {Chen}}, \bibinfo {author} {\bibfnamefont {L.}~\bibnamefont {Lei}},\ and\ \bibinfo {author} {\bibfnamefont {S.}~\bibnamefont {Shi}},\ }\href@noop {} {\bibfield  {journal} {\bibinfo  {journal} {Langmuir}\ }\textbf {\bibinfo {volume} {39}},\ \bibinfo {pages} {2408} (\bibinfo {year} {2023})}\BibitemShut {NoStop}%
\bibitem [{\citenamefont {Truzzolillo}\ \emph {et~al.}(2018)\citenamefont {Truzzolillo}, \citenamefont {Sennato}, \citenamefont {Sarti}, \citenamefont {Casciardi}, \citenamefont {Bazzoni},\ and\ \citenamefont {Bordi}}]{truzzolillo2018}%
  \BibitemOpen
  \bibfield  {author} {\bibinfo {author} {\bibfnamefont {D.}~\bibnamefont {Truzzolillo}}, \bibinfo {author} {\bibfnamefont {S.}~\bibnamefont {Sennato}}, \bibinfo {author} {\bibfnamefont {S.}~\bibnamefont {Sarti}}, \bibinfo {author} {\bibfnamefont {S.}~\bibnamefont {Casciardi}}, \bibinfo {author} {\bibfnamefont {C.}~\bibnamefont {Bazzoni}},\ and\ \bibinfo {author} {\bibfnamefont {F.}~\bibnamefont {Bordi}},\ }\href {https://doi.org/10.1039/c7sm02357j} {\bibfield  {journal} {\bibinfo  {journal} {Soft Matter}\ }\textbf {\bibinfo {volume} {14}},\ \bibinfo {pages} {4110} (\bibinfo {year} {2018})}\BibitemShut {NoStop}%
\bibitem [{\citenamefont {Truzzolillo}\ \emph {et~al.}(2015)\citenamefont {Truzzolillo}, \citenamefont {Roger}, \citenamefont {Dupas}, \citenamefont {Mora},\ and\ \citenamefont {Cipelletti}}]{truzzolillo2015}%
  \BibitemOpen
  \bibfield  {author} {\bibinfo {author} {\bibfnamefont {D.}~\bibnamefont {Truzzolillo}}, \bibinfo {author} {\bibfnamefont {V.}~\bibnamefont {Roger}}, \bibinfo {author} {\bibfnamefont {C.}~\bibnamefont {Dupas}}, \bibinfo {author} {\bibfnamefont {S.}~\bibnamefont {Mora}},\ and\ \bibinfo {author} {\bibfnamefont {L.}~\bibnamefont {Cipelletti}},\ }\href {https://doi.org/10.1088/0953-8984/27/19/194103} {\bibfield  {journal} {\bibinfo  {journal} {Journal of Physics: Condensed Matter}\ }\textbf {\bibinfo {volume} {27}},\ \bibinfo {pages} {194103} (\bibinfo {year} {2015})}\BibitemShut {NoStop}%
\bibitem [{\citenamefont {Amendola}\ \emph {et~al.}(2017)\citenamefont {Amendola}, \citenamefont {Pilot}, \citenamefont {Frasconi}, \citenamefont {Marag{\`o}},\ and\ \citenamefont {Iat{\`\i}}}]{amendola2017}%
  \BibitemOpen
  \bibfield  {author} {\bibinfo {author} {\bibfnamefont {V.}~\bibnamefont {Amendola}}, \bibinfo {author} {\bibfnamefont {R.}~\bibnamefont {Pilot}}, \bibinfo {author} {\bibfnamefont {M.}~\bibnamefont {Frasconi}}, \bibinfo {author} {\bibfnamefont {O.~M.}\ \bibnamefont {Marag{\`o}}},\ and\ \bibinfo {author} {\bibfnamefont {M.~A.}\ \bibnamefont {Iat{\`\i}}},\ }\href {https://doi.org/10.1088/1361-648x/aa60f3} {\bibfield  {journal} {\bibinfo  {journal} {Journal of Physics: Condensed Matter}\ }\textbf {\bibinfo {volume} {29}},\ \bibinfo {pages} {203002} (\bibinfo {year} {2017})}\BibitemShut {NoStop}%
\bibitem [{\citenamefont {Zygadlo}\ \emph {et~al.}(2024)\citenamefont {Zygadlo}, \citenamefont {Liu}, \citenamefont {Bernardo}, \citenamefont {Ai}, \citenamefont {Nieh},\ and\ \citenamefont {Hanson}}]{zygadlo2024}%
  \BibitemOpen
  \bibfield  {author} {\bibinfo {author} {\bibfnamefont {K.}~\bibnamefont {Zygadlo}}, \bibinfo {author} {\bibfnamefont {C.-H.}\ \bibnamefont {Liu}}, \bibinfo {author} {\bibfnamefont {E.~R.}\ \bibnamefont {Bernardo}}, \bibinfo {author} {\bibfnamefont {H.}~\bibnamefont {Ai}}, \bibinfo {author} {\bibfnamefont {M.-P.}\ \bibnamefont {Nieh}},\ and\ \bibinfo {author} {\bibfnamefont {L.~A.}\ \bibnamefont {Hanson}},\ }\href {https://doi.org/10.1039/d3na00758h} {\bibfield  {journal} {\bibinfo  {journal} {Nanoscale Advances}\ }\textbf {\bibinfo {volume} {6}},\ \bibinfo {pages} {146} (\bibinfo {year} {2024})}\BibitemShut {NoStop}%
\bibitem [{\citenamefont {Aili}\ \emph {et~al.}(2011)\citenamefont {Aili}, \citenamefont {Gryko}, \citenamefont {Sepulveda}, \citenamefont {Dick}, \citenamefont {Kirby}, \citenamefont {Heenan}, \citenamefont {Baltzer}, \citenamefont {Liedberg}, \citenamefont {Ryan},\ and\ \citenamefont {Stevens}}]{aili2011}%
  \BibitemOpen
  \bibfield  {author} {\bibinfo {author} {\bibfnamefont {D.}~\bibnamefont {Aili}}, \bibinfo {author} {\bibfnamefont {P.}~\bibnamefont {Gryko}}, \bibinfo {author} {\bibfnamefont {B.}~\bibnamefont {Sepulveda}}, \bibinfo {author} {\bibfnamefont {J.~A.}\ \bibnamefont {Dick}}, \bibinfo {author} {\bibfnamefont {N.}~\bibnamefont {Kirby}}, \bibinfo {author} {\bibfnamefont {R.}~\bibnamefont {Heenan}}, \bibinfo {author} {\bibfnamefont {L.}~\bibnamefont {Baltzer}}, \bibinfo {author} {\bibfnamefont {B.}~\bibnamefont {Liedberg}}, \bibinfo {author} {\bibfnamefont {M.~P.}\ \bibnamefont {Ryan}},\ and\ \bibinfo {author} {\bibfnamefont {M.~M.}\ \bibnamefont {Stevens}},\ }\href {https://doi.org/10.1021/nl203559s} {\bibfield  {journal} {\bibinfo  {journal} {Nano letters}\ }\textbf {\bibinfo {volume} {11}},\ \bibinfo {pages} {5564} (\bibinfo {year} {2011})}\BibitemShut {NoStop}%
\bibitem [{\citenamefont {Chowdhury}\ \emph {et~al.}(2004)\citenamefont {Chowdhury}, \citenamefont {Julian}, \citenamefont {Coates},\ and\ \citenamefont {Cot{\'e}}}]{chowdhury2004}%
  \BibitemOpen
  \bibfield  {author} {\bibinfo {author} {\bibfnamefont {M.~H.}\ \bibnamefont {Chowdhury}}, \bibinfo {author} {\bibfnamefont {A.~M.}\ \bibnamefont {Julian}}, \bibinfo {author} {\bibfnamefont {C.~J.}\ \bibnamefont {Coates}},\ and\ \bibinfo {author} {\bibfnamefont {G.~L.}\ \bibnamefont {Cot{\'e}}},\ }\href {https://doi.org/10.1117/1.1803847} {\bibfield  {journal} {\bibinfo  {journal} {Journal of Biomedical Optics}\ }\textbf {\bibinfo {volume} {9}},\ \bibinfo {pages} {1347} (\bibinfo {year} {2004})}\BibitemShut {NoStop}%
\bibitem [{\citenamefont {Guinier}\ and\ \citenamefont {Fournet}(1955)}]{guinier1955}%
  \BibitemOpen
  \bibfield  {author} {\bibinfo {author} {\bibfnamefont {A.}~\bibnamefont {Guinier}}\ and\ \bibinfo {author} {\bibfnamefont {G.}~\bibnamefont {Fournet}},\ }\href@noop {} {\bibinfo {title} {Small-angle scattering of x-rays}} (\bibinfo {year} {1955})\BibitemShut {NoStop}%
\bibitem [{\citenamefont {Grest}\ and\ \citenamefont {Kremer}(1986)}]{grest1986}%
  \BibitemOpen
  \bibfield  {author} {\bibinfo {author} {\bibfnamefont {G.~S.}\ \bibnamefont {Grest}}\ and\ \bibinfo {author} {\bibfnamefont {K.}~\bibnamefont {Kremer}},\ }\href {https://doi.org/10.1103/physreva.33.3628} {\bibfield  {journal} {\bibinfo  {journal} {Physical Review A}\ }\textbf {\bibinfo {volume} {33}},\ \bibinfo {pages} {3628} (\bibinfo {year} {1986})}\BibitemShut {NoStop}%
\bibitem [{\citenamefont {Kremer}\ and\ \citenamefont {Grest}(1990)}]{kremer1990}%
  \BibitemOpen
  \bibfield  {author} {\bibinfo {author} {\bibfnamefont {K.}~\bibnamefont {Kremer}}\ and\ \bibinfo {author} {\bibfnamefont {G.~S.}\ \bibnamefont {Grest}},\ }\href {https://doi.org/10.1063/1.458541} {\bibfield  {journal} {\bibinfo  {journal} {The Journal of Chemical Physics}\ }\textbf {\bibinfo {volume} {92}},\ \bibinfo {pages} {5057} (\bibinfo {year} {1990})}\BibitemShut {NoStop}%
\bibitem [{\citenamefont {Moreno}\ and\ \citenamefont {Verso}(2018)}]{moreno2018}%
  \BibitemOpen
  \bibfield  {author} {\bibinfo {author} {\bibfnamefont {A.~J.}\ \bibnamefont {Moreno}}\ and\ \bibinfo {author} {\bibfnamefont {F.~L.}\ \bibnamefont {Verso}},\ }\href {https://doi.org/10.1039/c8sm01407h} {\bibfield  {journal} {\bibinfo  {journal} {Soft Matter}\ }\textbf {\bibinfo {volume} {14}},\ \bibinfo {pages} {7083} (\bibinfo {year} {2018})}\BibitemShut {NoStop}%
\bibitem [{\citenamefont {Deserno}\ and\ \citenamefont {Holm}(1998)}]{deserno1998}%
  \BibitemOpen
  \bibfield  {author} {\bibinfo {author} {\bibfnamefont {M.}~\bibnamefont {Deserno}}\ and\ \bibinfo {author} {\bibfnamefont {C.}~\bibnamefont {Holm}},\ }\href {https://doi.org/10.1063/1.477414} {\bibfield  {journal} {\bibinfo  {journal} {The Journal of Chemical Physics}\ }\textbf {\bibinfo {volume} {109}},\ \bibinfo {pages} {7678} (\bibinfo {year} {1998})}\BibitemShut {NoStop}%
\bibitem [{\citenamefont {Del~Monte}\ \emph {et~al.}(2020)\citenamefont {Del~Monte}, \citenamefont {Camerin}, \citenamefont {Ninarello}, \citenamefont {Gnan}, \citenamefont {Rovigatti},\ and\ \citenamefont {Zaccarelli}}]{del2020}%
  \BibitemOpen
  \bibfield  {author} {\bibinfo {author} {\bibfnamefont {G.}~\bibnamefont {Del~Monte}}, \bibinfo {author} {\bibfnamefont {F.}~\bibnamefont {Camerin}}, \bibinfo {author} {\bibfnamefont {A.}~\bibnamefont {Ninarello}}, \bibinfo {author} {\bibfnamefont {N.}~\bibnamefont {Gnan}}, \bibinfo {author} {\bibfnamefont {L.}~\bibnamefont {Rovigatti}},\ and\ \bibinfo {author} {\bibfnamefont {E.}~\bibnamefont {Zaccarelli}},\ }\href@noop {} {\bibfield  {journal} {\bibinfo  {journal} {Journal of Physics: Condensed Matter}\ }\textbf {\bibinfo {volume} {33}},\ \bibinfo {pages} {084001} (\bibinfo {year} {2020})}\BibitemShut {NoStop}%
\bibitem [{\citenamefont {Del~Monte}\ \emph {et~al.}(2021)\citenamefont {Del~Monte}, \citenamefont {Truzzolillo}, \citenamefont {Camerin}, \citenamefont {Ninarello}, \citenamefont {Chauveau}, \citenamefont {Tavagnacco}, \citenamefont {Gnan}, \citenamefont {Rovigatti}, \citenamefont {Sennato},\ and\ \citenamefont {Zaccarelli}}]{del2021}%
  \BibitemOpen
  \bibfield  {author} {\bibinfo {author} {\bibfnamefont {G.}~\bibnamefont {Del~Monte}}, \bibinfo {author} {\bibfnamefont {D.}~\bibnamefont {Truzzolillo}}, \bibinfo {author} {\bibfnamefont {F.}~\bibnamefont {Camerin}}, \bibinfo {author} {\bibfnamefont {A.}~\bibnamefont {Ninarello}}, \bibinfo {author} {\bibfnamefont {E.}~\bibnamefont {Chauveau}}, \bibinfo {author} {\bibfnamefont {L.}~\bibnamefont {Tavagnacco}}, \bibinfo {author} {\bibfnamefont {N.}~\bibnamefont {Gnan}}, \bibinfo {author} {\bibfnamefont {L.}~\bibnamefont {Rovigatti}}, \bibinfo {author} {\bibfnamefont {S.}~\bibnamefont {Sennato}},\ and\ \bibinfo {author} {\bibfnamefont {E.}~\bibnamefont {Zaccarelli}},\ }\href@noop {} {\bibfield  {journal} {\bibinfo  {journal} {Proceedings of the National Academy of Sciences}\ }\textbf {\bibinfo {volume} {118}},\ \bibinfo {pages} {e2109560118} (\bibinfo {year} {2021})}\BibitemShut {NoStop}%
\bibitem [{\citenamefont {Plimpton}(1995)}]{plimpton1995}%
  \BibitemOpen
  \bibfield  {author} {\bibinfo {author} {\bibfnamefont {S.}~\bibnamefont {Plimpton}},\ }\href {https://doi.org/10.1006/jcph.1995.1039} {\bibfield  {journal} {\bibinfo  {journal} {Journal of computational physics}\ }\textbf {\bibinfo {volume} {117}},\ \bibinfo {pages} {1} (\bibinfo {year} {1995})}\BibitemShut {NoStop}%
\bibitem [{\citenamefont {Juba}\ \emph {et~al.}(2017)\citenamefont {Juba}, \citenamefont {Audus}, \citenamefont {Mascagni}, \citenamefont {Douglas},\ and\ \citenamefont {Keyrouz}}]{zeno}%
  \BibitemOpen
  \bibfield  {author} {\bibinfo {author} {\bibfnamefont {D.}~\bibnamefont {Juba}}, \bibinfo {author} {\bibfnamefont {D.~J.}\ \bibnamefont {Audus}}, \bibinfo {author} {\bibfnamefont {M.}~\bibnamefont {Mascagni}}, \bibinfo {author} {\bibfnamefont {J.~F.}\ \bibnamefont {Douglas}},\ and\ \bibinfo {author} {\bibfnamefont {W.}~\bibnamefont {Keyrouz}},\ }\bibfield  {journal} {\bibinfo  {journal} {Journal of Research of National Institute of Standards and Technology}\ }\textbf {\bibinfo {volume} {122}},\ \href {https://doi.org/10.6028/jres.122.020} {10.6028/jres.122.020} (\bibinfo {year} {2017})\BibitemShut {NoStop}%
\bibitem [{COM()}]{COMSOL}%
  \BibitemOpen
  \href@noop {} {\bibinfo {title} {{COMSOL} {M}ultiphysics\textsuperscript{\textregistered} v. 6.2.}},\ \bibinfo {howpublished} {\url{https://www.comsol.com}},\ \bibinfo {note} {{COMSOL AB, Stockholm, Sweden}}\BibitemShut {NoStop}%
\bibitem [{\citenamefont {Yushanov}\ \emph {et~al.}(2013)\citenamefont {Yushanov}, \citenamefont {Crompton},\ and\ \citenamefont {Koppenhoefer}}]{Yushanov2013}%
  \BibitemOpen
  \bibfield  {author} {\bibinfo {author} {\bibfnamefont {S.}~\bibnamefont {Yushanov}}, \bibinfo {author} {\bibfnamefont {J.~S.}\ \bibnamefont {Crompton}},\ and\ \bibinfo {author} {\bibfnamefont {K.~C.}\ \bibnamefont {Koppenhoefer}},\ }in\ \href@noop {} {\emph {\bibinfo {booktitle} {Proceedings of the COMSOL Conference}}},\ Vol.\ \bibinfo {volume} {116}\ (\bibinfo {organization} {Boston},\ \bibinfo {year} {2013})\ pp.\ \bibinfo {pages} {1--7}\BibitemShut {NoStop}%
\bibitem [{\citenamefont {Brasse}\ \emph {et~al.}(2018)\citenamefont {Brasse}, \citenamefont {M\"uller}, \citenamefont {Karg}, \citenamefont {Kuttner}, \citenamefont {K\"onig},\ and\ \citenamefont {Fery}}]{Brasse2018}%
  \BibitemOpen
  \bibfield  {author} {\bibinfo {author} {\bibfnamefont {Y.}~\bibnamefont {Brasse}}, \bibinfo {author} {\bibfnamefont {M.~B.}\ \bibnamefont {M\"uller}}, \bibinfo {author} {\bibfnamefont {M.}~\bibnamefont {Karg}}, \bibinfo {author} {\bibfnamefont {C.}~\bibnamefont {Kuttner}}, \bibinfo {author} {\bibfnamefont {T.~A.}\ \bibnamefont {K\"onig}},\ and\ \bibinfo {author} {\bibfnamefont {A.}~\bibnamefont {Fery}},\ }\href@noop {} {\bibfield  {journal} {\bibinfo  {journal} {ACS applied materials \& interfaces}\ }\textbf {\bibinfo {volume} {10}},\ \bibinfo {pages} {3133} (\bibinfo {year} {2018})}\BibitemShut {NoStop}%
\bibitem [{\citenamefont {Camerin}\ \emph {et~al.}(2018)\citenamefont {Camerin}, \citenamefont {Gnan}, \citenamefont {Rovigatti},\ and\ \citenamefont {Zaccarelli}}]{Camerin2018}%
  \BibitemOpen
  \bibfield  {author} {\bibinfo {author} {\bibfnamefont {F.}~\bibnamefont {Camerin}}, \bibinfo {author} {\bibfnamefont {N.}~\bibnamefont {Gnan}}, \bibinfo {author} {\bibfnamefont {L.}~\bibnamefont {Rovigatti}},\ and\ \bibinfo {author} {\bibfnamefont {E.}~\bibnamefont {Zaccarelli}},\ }\href@noop {} {\bibfield  {journal} {\bibinfo  {journal} {Scientific reports}\ }\textbf {\bibinfo {volume} {8}},\ \bibinfo {pages} {14426} (\bibinfo {year} {2018})}\BibitemShut {NoStop}%
\bibitem [{\citenamefont {Bischofberger}\ and\ \citenamefont {Trappe}(2015)}]{Bischofberger2015}%
  \BibitemOpen
  \bibfield  {author} {\bibinfo {author} {\bibfnamefont {I.}~\bibnamefont {Bischofberger}}\ and\ \bibinfo {author} {\bibfnamefont {V.}~\bibnamefont {Trappe}},\ }\href@noop {} {\bibfield  {journal} {\bibinfo  {journal} {Scientific reports}\ }\textbf {\bibinfo {volume} {5}},\ \bibinfo {pages} {15520} (\bibinfo {year} {2015})}\BibitemShut {NoStop}%
\bibitem [{\citenamefont {Raki{\'c}}\ \emph {et~al.}(1998)\citenamefont {Raki{\'c}}, \citenamefont {Djuri{\v{s}}i{\'c}}, \citenamefont {Elazar},\ and\ \citenamefont {Majewski}}]{Rakic1998}%
  \BibitemOpen
  \bibfield  {author} {\bibinfo {author} {\bibfnamefont {A.~D.}\ \bibnamefont {Raki{\'c}}}, \bibinfo {author} {\bibfnamefont {A.~B.}\ \bibnamefont {Djuri{\v{s}}i{\'c}}}, \bibinfo {author} {\bibfnamefont {J.~M.}\ \bibnamefont {Elazar}},\ and\ \bibinfo {author} {\bibfnamefont {M.~L.}\ \bibnamefont {Majewski}},\ }\href@noop {} {\bibfield  {journal} {\bibinfo  {journal} {Applied optics}\ }\textbf {\bibinfo {volume} {37}},\ \bibinfo {pages} {5271} (\bibinfo {year} {1998})}\BibitemShut {NoStop}%
\bibitem [{\citenamefont {Laurens}\ and\ \citenamefont {Oughstun}(1999)}]{Laurens1999}%
  \BibitemOpen
  \bibfield  {author} {\bibinfo {author} {\bibfnamefont {J.~E.}\ \bibnamefont {Laurens}}\ and\ \bibinfo {author} {\bibfnamefont {K.~E.}\ \bibnamefont {Oughstun}},\ }in\ \href@noop {} {\emph {\bibinfo {booktitle} {Ultra-Wideband Short-Pulse Electromagnetics 4 (IEEE Cat. No. 98EX112)}}}\ (\bibinfo {organization} {IEEE},\ \bibinfo {year} {1999})\ pp.\ \bibinfo {pages} {243--264}\BibitemShut {NoStop}%
\end{thebibliography}%

\end{document}